\documentclass[authoryear]{elsarticle} 
\usepackage[hyphens]{url}

  \journal{Fisheries Research} 

\usepackage{lineno} 
\providecommand{\tightlist}{%
  \setlength{\itemsep}{0pt}\setlength{\parskip}{0pt}}

\usepackage{graphicx}
\usepackage{booktabs} 

\usepackage[T1]{fontenc}
\usepackage{lmodern}
\usepackage{amssymb,amsmath}
\usepackage{ifxetex,ifluatex}
\usepackage{fixltx2e} 
\IfFileExists{upquote.sty}{\usepackage{upquote}}{}
\ifnum 0\ifxetex 1\fi\ifluatex 1\fi=0 
  \usepackage[utf8]{inputenc}
\else 
  \usepackage{fontspec}
  \ifxetex
    \usepackage{xltxtra,xunicode}
  \fi
  \defaultfontfeatures{Mapping=tex-text,Scale=MatchLowercase}
  
\fi
\IfFileExists{microtype.sty}{\usepackage{microtype}}{}
\usepackage[letterpaper]{geometry}
\usepackage{natbib}
\usepackage{longtable}
\ifxetex
  \usepackage[setpagesize=false, 
              unicode=false, 
              xetex]{hyperref}
\else
  \usepackage[unicode=true]{hyperref}
\fi
\hypersetup{breaklinks=true,
            bookmarks=true,
            pdfauthor={},
            pdftitle={Hierarchical stock assessment methods improve management performance in multi-species, data-limited fisheries.},
            colorlinks=false,
            urlcolor=blue,
            linkcolor=magenta,
            pdfborder={0 0 0}}
\newlength{\cslhangindent}
  {}%
  {\par}

\usepackage{lscape}
\usepackage{pdflscape}
\newcommand{\blandscape}{\begin{landscape}}
\newcommand{\elandscape}{\end{landscape}}

\newpageafter{abstract}
\begin{document}
\begin{frontmatter}

  \title{Hierarchical stock assessment methods improve management performance in multi-species, data-limited fisheries.}
    \author[a]{Samuel D. N. Johnson\corref{1}}
   \ead{samuelj@sfu.ca} 
    \author[a]{Sean P. Cox}
   \ead{spcox@sfu.ca} 
      \address[a]{School of Resource and Environmental Management, Simon Fraser University, 8888 University Drive, BC, V5K 1S6, Canada}
      \cortext[1]{Corresponding author}
  
  \begin{abstract}
  Managers of multi-species fisheries aim to balance harvest of multiple
  interacting species that vary in abundance and productivity. Technical
  interactions are a defining characteristic of multi-species fisheries,
  and are often ignored when setting catch limits or target exploitation
  rates, despite their effect on the ability to balance multi-species
  harvest. Balancing harvest is more difficult when stock assessments are
  impossible to perform due to data-limitations, involving \emph{inter alia}
  short time series of fishery independent observations, poor catch
  monitoring, limited biological data, and spatial mismatching of
  harvest decisions to system structure. In this paper, we asked whether
  using hierarchical, data-limited stock assessment methods to set catch
  limits improved management performance in data-limited, multi-species
  fisheries. Management performance of five alternative
  stock assessment methods was evaluated by using them to set harvest
  levels targeting multi-species maximum yield in a multi-species flatfish
  fishery, including single-species and hierarchical multi-species
  models, and methods that pooled data across species and spatial
  strata, with catch outcomes of each method under three data scenarios
  compared to catch under an omniscient manager simulation.
  Operating models included technical interactions between species
  intended to produce choke effects often observed in output
  controlled multi-species fisheries. Hierarchical multi-species models
  outperformed all other methods under data-poor and data-moderate
  scenarios, and outperformed single-species models under the data-rich
  scenario. Hierarchical models were least sensitive to prior precision,
  sometimes improving in performance when prior precision was reduced.
  Choke effects were found to both positive and negative effects,
  sometimes leading to underfishing of non-choke species, but at other
  times preventing overfishing of non-choke species. We highlight the
  importance of including technical interactions in multi-species
  assessment models and management objectives, how choke species can
  indicate mismatches between management objectives and system dynamics,
  and recommend hierarchical multi-species models for multi-species
  fishery management systems.
  \end{abstract}
 \end{frontmatter}

\hypertarget{introduction}{%
\section{Introduction}\label{introduction}}

Managers of multi-species fisheries aim to balance harvest of multiple
interacting target and non-target species that vary in abundance and
productivity. Among-species variation in productivity implies
variation in single-species optimal harvest rates, and, therefore,
differential responses to exploitation. Single-species optimal harvest
rates (e.g., the harvest rate associated with maximum sustainable
yield) typically ignore both multi-species trophic interactions that
influence individual demographic rates \citep{gislason1999single, collie2001biological}, and technical interactions that make it virtually
impossible to simultaneously achieve the optimal harvest rate for all
species \citep{pikitch1987use}.

Technical interactions among species that co-occur in non-selective
fishing gear are a defining characteristic of multi-species fisheries
\citep{pikitch1987use, punt2002evaluation} and, therefore, play a central
role in multi-species fisheries management outcomes for individual
species \citep{ono2017management, kempf2016msy}. Catch limits set
for individual species without considering technical interactions
subsequently lead to sub-optimal fishery outcomes \citep{ono2017management, punt2011calculating, punt2020multispecies}. For example, under-utilization
of catch limits could occur when technically interacting quota species are
caught at different rates (i.e., catchability) by a common gear, leading
to a choke constraint in which one species quota is filled before the
others \citep{baudron2015adverse}. Choke constraints are generally
considered negative outcomes for multi-species fishery performance,
because they reduce harvester profitability as increasingly rare quota
for choke species may limit access to fishing grounds, as well as
driving quota costs above the landed value of the choke
species \citep{mortensen2018identifying}.

Setting catch limits for individual species in any fishery usually
requires an estimate of species abundance, which continues to be
a central challenge of fisheries stock assessment
\citep{hilborn1992quantitative, quinn2003ruminations, maunder2015contemporary},
especially when species data are insufficient for stock assessment models. Where such data limitations
exist, data pooling is sometimes used to extend stock assessments to
complexes of similar, interacting stocks of fish \citep{appeldoorn1996model}.
A couple of examples include pooling data for a single species across
multiple spatial strata when finer scale data are unavailable
or when fish are believed to move between areas at a sufficiently
high rate \citep{benson2015evaluating, punt2018estimates}, and pooling data
for multiple species of the same taxonomic group within an area
when data are insufficient for individual species or during development
of new fisheries \citep{demartini2019hazards}. Data-pooled estimates of
productivity represent means across the species complex; therefore the
general cost of such pooling approaches is that catch limits will tend to
overfish unproductive species and underfish productive ones
\citep{gaichas2012assembly}. Hierarchical stock assessment models may help reduce
these problems by representing a compromise between uninformative
single-species single-species approaches and problematic data-pooling
approaches. In particular, hierarchical models allow data sharing
between multiple species within a complex via hierarchical shrinkage
prior distributions on model parameters that represent biological
similarities or technical interactions \citep{thorson2015giants}. Shared priors
shrink species-specific parameters towards an estimated complex level
mean value, improving model convergence for data-poor species and basing
estimates more on available data than strong \emph{a priori} assumptions like
fixed (i.e.~known) parameters, identical parameters among species, or
strongly informative priors \citep{jiao2009hierarchical, jiao2011poor, punt2011among}.

Although hierarchical stock assessments produce better
estimates of species biomass and productivity than single-species
methods in data-limited contexts, it remains unclear whether
such improved statistical performance translates into better
management outcomes \citep{johnson2018evaluating}. On one hand, improving
assessment model statistical performance may sacrifice future
fishery benefits of adaptive learning \citep{walters1986adaptive},
as large assessment errors caused by poor quality data in the
present may generate high contrast data that are more informative
in the future. Alternatively, adaptive learning may fail to occur
in a reasonable time in the presence of large errors, leading
to stock collapse or harvesters divesting from the fishery because
of over- or under-exploitation. Improving the rate of adaptive
learning may be possible via hierarchical modeling of spatially
replicated groundfish stocks, because the shared priors take
lessons learned from responses to disturbances in one area
and use them to improve stock assessments in other areas
\citep{collie1991adaptive}.

In this paper, we investigated whether hierarchical
stock assessment models improved management performance
in a simulated multi-species, data-limited fishery. The
simulated fishery was modeled on a complex of Dover sole,
English sole, and southern rock sole - three species of
right-eyed flounders (\emph{Pleuronectidae spp.}) - Dover sole (\emph{Microstomus
pacificus}), English sole (\emph{Parophrys vetulus}), and southern rock
sol (\emph{Lepidopsetta bilineata}) - fished off the
coast of British Columbia, Canada, in three spatial management areas.
Closed-loop feedback simulation was used to estimate fishery
outcomes when catch limits were set based on estimates
of biomass from single-species, data-pooling, and hierarchical
state-space surplus production models under three data scenarios.
Assessment models were either fit to species specific data as
single-species or hierarchical multi-species models, or fit to data
pooled spatially across management units, pooled across species
within a spatial management unit, or totally aggregated across
both species and spatial management units. Management performance of
each assessment approach was measured as cumulative absolute loss in
catch, measured against optimal catch trajectories generated by an
omniscient manager, who could set annual effort to maximize total
multi-species/multi-stock complex yield given perfect knowledge of all
future recruitments \citep{walters1998evaluation, martell2008retrospective}.

\hypertarget{methods}{%
\section{Methods}\label{methods}}

\hypertarget{british-columbias-flatfish-fishery}{%
\subsection{British Columbia's flatfish fishery}\label{british-columbias-flatfish-fishery}}

British Columbia's multi-species complex of right-eyed flounders is
a technically interacting group of flatfishes managed across the whole
BC coast. Although there are several right-eyed flounders in BC waters,
we focus on Dover sole, English sole, and Rock sole. Taken together,
these species comprise a multi-stock complex (DER complex), managed
as part of the BC multi-species groundfish fishery.

The DER complex is managed in three spatially distinct stock
areas (Figure 1) \citep{PRIFMP2015}. From north to south, the first
stock area - Hecate Strait/Haida Gwaii (HSHG) - corresponds to
BC groundfish management area 5CDE, extending from
Dixon Entrance and north of Haida Gwaii, south through
Hecate Strait. The second stock area - Queen Charlotte Sound
(QCS) - corresponds to BC groundfish management area 5AB,
extending from the southern tip of Haida Gwaii to the northern
tip of Vancouver Island. Finally, the third area - West Coast of
Vancouver Island (WCVI) - corresponds to groundfish management
area 3CD, extending from the northern tip of Vancouver Island
south to Juan de Fuca Strait.

Each area had two commercial catch rate series and at least one
survey biomass index time series for each species (Table 1).
The two commercial fishery catch-per-unit-effort series span
1976 - 2016, split between a historical trawl fishery
(1976 - 1995) and a modern trawl fishery (1996 - 2016),
with the split corresponding to pre- and post-implementation of an
at-sea-observer program, respectively. The fishery independent trawl
survey biomass indices were the biennial Hecate Strait Assemblage
survey in the HSHG area (1984 - 2002), and the Groundfish Multi-species
Synoptic Survey, with separate biennial legs in all three stock
areas (2003 - 2016).

\hypertarget{closed-loop-feedback-simulation-framework}{%
\subsection{Closed-loop feedback simulation framework}\label{closed-loop-feedback-simulation-framework}}

Closed-loop simulation is often used to evaluate proposed
feedback management systems. In fisheries, closed-loop simulation evaluates
fishery management system components, such as stock assessment models or
harvest decision rules, by simulating repeated applications of these
components, along with propagating realistic errors in monitoring data,
stock assessment model outputs, and harvest advice
\citep{de-la-Mare1998Tidier-fisherie, cox2008practical, cox2013roles}. At the
end of the simulation, pred-defined performance metrics are used to
determine the relative performance of the system components being tested.

Our closed-loop simulation framework included a stochastic operating,
representing the stock and fishery dynamics, as well as an observation
model, and an assessment model component that estimated stock biomass
from simulated observations and fishery catches. The
operating model simulated population dynamics of a spatially
stratified, multi-species flatfish complex in response to a
multi-species trawl fishery in each of the three stock areas.
Although total fishing effort was not restricted across the
entire area, effort in each individual area was allocated
such that no species-/area-specific catch exceeded the
species-/area-specific total allowable catch (TAC). Within an area,
fishing effort was allowed to increase until at least one
species-/area-specific TAC was fully caught.

The simulation projected population dynamics for all nine stocks
forward in time for \(32\) years, with annual
simulated assessments, harvest decisions, and catch removed from
the population from 2017 - 2048. The following four steps summarise
the closed-loop simulation procedure for each projection year \(t\):

\begin{enumerate}
\def\labelenumi{\arabic{enumi}.}
\tightlist
\item
  Update stochastic population dynamics and
  generate new realized catch \(C_{s,p,t}(E_{p,t})\) in each
  area from effort (eq \ref{eq:FqE});
\item
  Generate new observation data \(I_{s,p,f,t}\)
  (eqs \ref{eq:startIdx}-\ref{eq:endIdx})
\item
  Apply an assessment model (defined below) to estimate the
  spawning biomass for the upcoming year \(\hat{B}_{s,p,t+1}\)
  (eq \ref{eq:amBio});
\item
  Apply a harvest rate to generate a total allowable catch
  \(TAC_{s,p,t+1}\) (eqs \ref{eq:startHR} - \ref{eq:endHR});
\item
  Allocate effort \(E_{p,t+1}\) to fully realise at least
  one TAC in each stock area (eq \ref{eq:maxE}).
\end{enumerate}

\hypertarget{operating-model}{%
\subsubsection{Operating Model}\label{operating-model}}

The operating model (OM) was a multi-species, multi-stock age-
and sex-structured population dynamics model (Appendix A).
Population life-history parameters for the operating
model were estimated by fitting a hierarchical age- and
sex-structured model to data from the real DER complex
(Johnson, Cox and Knowler, in Prep).

Fishing mortality for individual stocks was scaled to
commercial trawl effort via species-specific catchability
parameters, i.e.,
\begin{equation}
F_{s,p,t} = q^{F}_{s,p} \cdot E_{p,t},\label{eq:FqE}
\end{equation}
where \(F_{s,p,t}\) is the fishing mortality rate applied
to species \(s\) in stock-area \(p\) by fleet \(f\) in year
\(t\), and \(q^{F}_{s,p}\) is the commercial catchability
coefficient scaling trawl effort \(E_{p,t}\) in area \(p\)
to fishing mortality (Table 2) {[}Johnson and Cox, in prep{]}.

The above effort model implies a multi-species maximum yield for
the trawl fishery in each area, depending on individual species
productivity, commercial trawl catchability, and relative
catchabilities among species in the area \citep{pikitch1987use, punt2011calculating}, i.e.,
\begin{equation*}
MSY_{MS,p} = \max_{E} Y_{MS} 
    \left( E, q^{F}_{1,p}, q^{F}_{2,p}, q^{F}_{3,p}, Q_p \right),
\end{equation*}
where \(E\) is the total commercial trawl fishing effort, \(q_{s,p}\)
is the commerical trawl catchability coefficient scaling fishing
effort to fishing mortality for species \(s\)
in area \(p\), and \(Q_p\) is the set of life-history parameters for
all DER complex species in area \(p\). The function \(Y_{MS}\) is the
total multi-species yield in area \(p\) as a function of fishing
effort \(E_p\) (Figure 2, heavy black lines), defined as the sum of
species-specific equilibrium yields
\begin{equation*}
Y_{MS,p} \left(E_p, q^{F}_{1,p}, q^{F}_{2,p}, q^{F}_{3,p}, Q_p \right)
 = \sum^3_{s=1} Y_{SS,s,p} (E_p, q_{s,p}, Q_{s,p} ),
\end{equation*}
where \(Q_{s,p}\) is the subset of \(Q_p\) containing life history
parameters for species \(s\) in area \(p\), and the function \(Y_{SS,s,p}\)
is the traditional single-species yield curve expressed as a
function of effort rather than fishing mortality (Figure 2,
coloured lines).

The level of fishing effort producing multi-species maximum
yield \(MSY_{MS,p}\) in area \(p\) is defined as
\begin{equation*}
E_{MSY,p} = {\mathop{\mathrm{arg\,max}}}_{E} Y_{MS} 
    \left( E, q^{F}_{1,p}, q^{F}_{2,p}, q^{F}_{3,p}, Q_p \right),
\end{equation*}
with arguments defined above (Figure 2, vertical dashed line).
The fishing effort level \(E_{MSY,p}\) has an associated equilibrium
biomass \(B_{MSY,MS,s,p}\) and yield
\begin{equation*}
MSY_{MS,s,p} = Y_{SS,s,p} (E_{MSY,p}, q^{F}_{s,p}, Q_{s,p} )
\end{equation*}
for each species \(s\) (Figure 2, three lower horizontal dashed
lines showing individual species yield), which differ from the
single-species \(B_{MSY}\) and \(MSY\) when commercial catchability
scalars differ between species.

For simulated fishing in the projection time period, we allocated
maximum fishing effort to each area, defined as the effort required
to fully utilize the TAC of at least one species, while never
exceeding the TAC of any one of the three species, i.e.,
\begin{equation}\label{eq:maxE}
E_{p,t+1} = \max \{ E ~|~ C_{s,p,t+1}(E) \leq TAC_{s,p,t+1} \forall s \},
\end{equation}
where \(C_{s,p,t+1}(E)\) is the catch of species \(s\) when effort \(E\)
is applied in area \(p\) in year \(t+1\) (the method for determining
TACs is explained below). We used maximum effort instead of an
explicit effort dynamics model because the former was adequate for
understanding the relative management performance of the assessment
models that we tested. Furthermore, the maximum effort model may be
simplified over reality, but it did an adequate job of capturing
choke effects observed in the real BC groundfish fishery system.

\hypertarget{surplus-production-stock-assessment-models}{%
\subsubsection{Surplus production stock assessment models}\label{surplus-production-stock-assessment-models}}

At each time step \(t\), simulated annual assessments were used to
estimate the expected future spawning biomass estimate \(\hat{B}_{t+1}\)
via a state-space Schaefer production model \citep{schaefer1957some, punt2002evaluation}, modified to better approximate the
biomass-yield relationship underlying the age-/sex-structured
operating model \citep{pella1969generalized, winker2018jabba}.
We extended the \citet{johnson2018evaluating} hierarchical state-space
model to fit a multi-species, spatially stratified complex,
as well as fit a single-stock model to data from individual
or data-pooled stocks, via the biomass equation
\begin{equation}\label{eq:amBio}
B_{s,p,t+1} = \left[ B_{s,p,t} + U_{MSY,s,p} \cdot \frac{m_{s,p}}{m_{s,p}-1} 
                  \cdot B_{s,p,t} \cdot \left(1 - \frac{1}{m_{s,p}} \cdot \left( \frac{B_{s,p,t}}{B_{MSY,s,p}} \right)^{m_{s,p} - 1}\right) - C_{s,p,t}\right] e^{\zeta_{s,p,t}},
\end{equation}
where \(B_{MSY}\) (optimal biomass) and \(U_{MSY}\) (optimal harvest
rate) are the leading biological model parameters, \(m_{s,p}\)
is the Pella-Tomlinson parameter controlling skew in the
biomass/yield relationship (derived from operating
model yield curve), and \(\zeta_{s,p,t}\) are annual
process error deviations.

In total, we defined the following five potential assessment
model configurations:

\begin{enumerate}
\def\labelenumi{\arabic{enumi}.}
\tightlist
\item
  Total Aggregation (1 stock);
\item
  Species Pooling (3 stocks, independent);
\item
  Spatial Pooling (3 species, independent);
\item
  Single-stock (9 stocks, independent);
\item
  Hierarchical Multi-stock (9 stocks, sharing information);
\end{enumerate}

\noindent where the number of management units is shown in
parentheses (delete subscripts in eq. 7 for species or stock-area as
appropriate, e.g., Spatial Pooling models have no \(p\) subscript).

Prior distributions on optimal biomass \(B_{MSY,s,p}\), optimal
harvest rate \(U_{MSY,s,p}\), catchability \(q_{s,p,f}\), and
process error deviations \(\zeta_{s,p,t}\) were defined for each
assessment model (Table 3). For all non-hierarchical models, prior mean values
were derived from the operating model biomass/yield relationship,
with prior mean \(B_{MSY}\) values
\begin{align*}
\mu_{B_{MSY,s,p}} &=  B_{MSY,s,p}, \\
\mu_{B_{MSY,s}}   &= \sum_{p} B_{MSY,s,p}, \\
\mu_{B_{MSY,p}}   &= \sum_{s} B_{MSY,s,p}, \\
\mu_{B_{MSY}}     &= \sum_{s,p} B_{MSY,s,p},
\end{align*}
for single-stock, spatial pooled, species pooled, and totally aggregated
configurations, where \(B_{MSY,s,p}\) is the equilibrium biomass at which
single-species maximum yield is achieved. Similarly, log-normal prior
mean \(\log U_{MSY}\) values were given by
\begin{align*}
\mu_{\log U_{MSY,s,p}} &= \log (MSY_{s,p} / B_{MSY,s,p}), \\
\mu_{\log U_{MSY,s}}   &= \log( \sum_{p} MSY_{s,p} / \sum_{p} B_{MSY,s,p} ), \\
\mu_{\log U_{MSY,p}}   &= \log( \sum_{s} MSY_{s,p} / \sum_{s} B_{MSY,s,p}), \\
\mu_{\log U_{MSY}}     &= \log( \sum_{s,p} MSY_{s,p} / \sum_{s,p} B_{MSY,s,p}),
\end{align*}
for single-stock, spatial-pooled, species-pooled, and totally aggregated
configurations, respectively, where \(MSY_{s,p}\) is the single-species
maximum yield. Log-normal survey biomass and commercial CPUE catchability
had prior means set to OM values for the single-stock model, and prior means
that were averaged OM values over the sources of pooled data under the
data-pooled models. Prior standard deviations for catchability were set to
\(s_f = 1.0\) for commercial CPUE indices in all models, \(s_f = 1.0\) for survey
biomass indices in data-pooled models, and \(s_f = 0.5\) for survey indices
in the single-stock model. Finally, process error prior standard deviations
were fixed at \(0.05\) for all stocks, species, and model configurations.

The hierarchical multi-stock model and the single-stock model
used the same priors for \(B_{MSY}\) and catchability
for the commercial CPUE and Hecate Strait Assemblage
survey biomass indices. Log-normal hierarchical shrinkage
priors were applied to area-specific Synoptic survey catchabilities
\(q_{s,p,f}\) within each species, and in two levels to \(U_{MSY,s,p}\)
(as a proxy for intrinsic growth rates) across the whole complex.
Both hierarchical prior distributions had fixed standard
deviation parameters \(\sigma_{U_{MSY}} = \tau_q = 0.05\)
for differences between stocks within a species for
catchability \(q_{s,p,f}\) and \(U_{MSY}\), and the same
standard deviation for the log-normal prior on species average
\(\overline{U}_{MSY,s}\) values. Hierarchical species mean
catchability for the Synoptic survey \(\overline{q}_{s,f}\)
had a log-normal hyperprior with mean \(m_{\log q_{s,f}}\) and
standard deviation \(s_f = .5\), and DER complex mean
\(\overline{U}_{MSY}\) had a log-normal hyperprior with
mean \(m_{\log U_{MSY}}\) and standard devation of
\(s_{U_{MSY}} = .4\).

\hypertarget{data-generation-for-assessment-models}{%
\subsubsection{Data generation for assessment models}\label{data-generation-for-assessment-models}}

Time series of catch and biomass indices were simulated in
the historical and projection periods for fitting assessment
models. Simulated data matched the time periods of the real
DER complex biomass index data sources, with the exception of
the Synoptic trawl survey, which was also simulated in the projection,
period (Table 1). Observation error precision for each data source
is given in Table A.1 (Appendix A).

Biomass indices for individual stocks depended on the nature of
the index. Assemblage and Synoptic Survey biomass indices were
defined as trawlable biomass
\begin{equation}\label{eq:startIdx}
\bar{I}_{s,p,f,t} = q_{s,p,f} \cdot B_{s,p,f,t},
\end{equation}
where \(\bar{I}_{s,p,f,t}\) is the index without observation error,
\(q_{s,p,f}\) is the survey trawl efficiency for species \(s\)
and stock-area \(p\), and \(B_{s,p,f,t}\) is the biomass of
species \(s\) in stock-area \(p\) vulnerable to survey \(f\) in
year \(t\). Commercial CPUE indices were defined as
\begin{equation}
\bar{I}_{s,p,f,t} = \frac{C_{s,p,f,t}}{E_{p,f,t}},
\end{equation}
where \(C_{s,p,f,t}\) was commercial catch by fleet \(f\) of
species \(s\) from area \(p\) at time \(t\), and \(E_{p,f,t}\) was
commercial fishing effort in area \(p\) from fleet \(f\) at
time \(t\), with both catch and effort required to be positive.

The method for generating pooled catch and biomass index data depended on
the data type. Catch data were pooled by summation, and the index data were
pooled according to the stock-specific definition above. For Assemblage
and Synoptic survey biomass indices, pooled indices without
observation error were defined as
\begin{align} 
\bar{I}^{pooled}_{s,f,t}  & = \sum_{p} \mathcal{I}(I_{s,p,f,t} > 0) \cdot q_{s,p,f} \cdot B_{s,p,f,t} \\
\bar{I}^{pooled}_{p,f,t}  & = \sum_{s} \mathcal{I}(I_{s,p,f,t} > 0) \cdot q_{s,p,f} \cdot B_{s,p,f,t} \\
\bar{I}^{pooled}_{f,t}    & = \sum_{s,p} \mathcal{I}(I_{s,p,f,t} > 0) \cdot q_{s,p,f} \cdot B_{s,p,f,t}
\end{align}
where \(\bar{I}^{pooled}_{s,f,t}\) is the spatially pooled index for species
\(s\), \(\bar{I}^{pooled}_{p,f,t}\) is a species pooled index for area \(p\),
\(\bar{I}^{pooled}_{f,t}\) is the totally aggregated index (all without
error), and \(\mathcal{I}(I_{s,p,f,t} > 0)\) is the indicator function that
takes value \(1\) when survey leg for fleet \(f\) in area \(p\) was running in
year \(t\). Pooled commercial CPUE indices were simulated as
\begin{align} 
\bar{I}^{pooled}_{s,f,t}  & = \frac{\sum_{p} C_{s,p,f,t}}{\sum_{p} E_{p,f,t} } \\
\bar{I}^{pooled}_{p,f,t}  & = \frac{\sum_{s} C_{s,p,f,t}}{ E_{p,f,t} } \\
\bar{I}^{pooled}_{f,t}    & = \frac{\sum_{s,p} C_{s,p,f,t}}{ \sum_{p} E_{p,f,t} } \label{eq:endIdx}
\end{align}
where \(C_{s,p,f,t}\) is catch, and \(E_{p,f,t}\) is commercial trawl
effort, with subscripts as defined above.

\hypertarget{target-harvest-rates-and-total-allowable-catch}{%
\subsubsection{Target harvest rates and total allowable catch}\label{target-harvest-rates-and-total-allowable-catch}}

Simulated harvest decision rules applied a constant target harvest
rate to generate TACs from one-year ahead biomass forecasts obtained
from each assessment model, i.e.,
\begin{equation}\label{eq:startHR}
TAC'_{s,p,t+1} =  U_{s,p} \cdot \hat{B}_{s,p,t+1},
\end{equation}
where \(U_{s,p}\) is the target harvest rate for species \(s\) in
stock-area \(p\), and \(\hat{B}_{s,p,t+1}\) is the year \(t+1\) biomass forecast
from the assessment model. Target harvest rates were derived
from the multi-species maximum yield relationship via
\begin{equation}
U_{s,p} = \frac{MSY_{MS,s,p}}{B_{MSY,MS,s,p}},
\end{equation}
where \(MSY_{MS,s,p}\) and \(B_{MSY,MS,s,p}\) are
the yield and spawning biomass, respectively,
associated with \(E_{MSY,MS,p}\) for species \(s\)
in area \(p\).

Finally, inter-annual increases in TAC were limited to 20\% for all
individual stocks as a practical precautionary measure, i.e.,
\begin{equation}
TAC_{s,p,t+1} = \min \{ TAC'_{s,p,t+1},1.2 * TAC_{s,p,t}\},
\end{equation}
where \(TAC'_{s,p,t+1}\) is the proposed TAC determined above, and
\(TAC_{s,p,t}\) is the previous year's TAC. The limited increase
reduces overfishing as a result of optimistic assessment errors,
potentially requiring several years of increases to realise target
harvest rates as set by assessment model estimates. Conversely, no
limit was applied to TAC reductions, so that steep declines in biomass
forecasts would lead to steep declines in catch. Note that this
constraint on inter-annual TAC changes is reflected as a constraint
on inter-annual changes in effort in the objective function used in
omniscient manager solutions described below.

Pooled TACs were set analogously to the stock-specific case above, with
pooled target harvest rates applied to biomass projections from pooled
assessments. For a spatially pooled assessment of species \(s\), we defined
the spatially pooled target harvest rate as
\begin{equation}
U_{s} = \frac{\sum_p MSY_{MS,s,p}}{\sum_p B_{MSY,MS,s,p}},
\end{equation}
where the notation is as defined above. Species pooled harvest rates and
total aggregation harvest rates were defined analogously.

Pooled TACs were split within an area or across spatial strata proportional
to Synoptic trawl survey indices for the individual stocks. For example, if
the TAC for area \(p\) is set by a species pooled assessment, then the
proposed TAC for species \(s\) is defined as
\begin{equation}\label{eq:endHR}
TAC'_{s,p,t+1} = \frac{\bar{I}_{s,p,2}}{\sum_{s'} \bar{I}_{s',p,2}} TAC_{p,t+1},
\end{equation}
where \(\bar{I}_{s,p,2}\) is the 2-year running average of individual biomass
indices from the Synoptic survey for species \(s\) in area \(p\). The 2-year
average is used because the synoptic survey alternates the legs each
year, so some individual stock indices are missing.

\hypertarget{simulation-experiments-and-performance}{%
\subsection{Simulation experiments and performance}\label{simulation-experiments-and-performance}}

We ran a total of \(15\) simulation experiments comprising five assessment
models and three data quality scenarios. Simulations integrated over the
stochastic processess by running at least 100 random replicates of each
combination, where each simulation was initialized with the same set of
random seeds to eliminate random effects among combinations
of assessments and data scenarios. Simulations stopped when they either
reached 100 replicates where simulated assessment models converged in at least 95\% of
projection time steps for all stocks, or when 140 random replicates were
attempted. We defined convergence as a positive definite Hessian matrix
and a maximum gradient component less than \(10^{-3}\) in absolute value.
A lower threshold of 95\% convergence was chosen given that fitting models
becomes more difficult as data quality is deliberately reduced, and a
simulated assessment can not always be tuned like a real assessment
performed by a real-life analyst. Any OM/assessment model/stock combinations that could
not reach 100 replicates meeting the 95\% convergence criterion over a
total of 140 attempts (approximately a 70\% success rate) were considered
non-significant and were indicated as such in the results. The operating
model was run for two Dover sole generations \citep[32 years;][]{seber1997estimation},
because this species had the longest generation time.

Operating model population dynamics and index data were
identical among replicates for each stock during the operating
model historical period, except for the last few
years near the end, where the operating model simulates
recruitment process errors because the conditioning assessment
was unable to estimate them. Simulated log-normal observation
and process errors in the projection were randomly drawn with
the same standard deviations as the errors used in the historical
period, and bias corrected so that asymptotic medians
matched their expected values, i.e.,
\begin{align}
I_{s,p,f,t} &= \bar{I}_{s,p,f,t} \cdot \exp( \tau_{s,p,f} \cdot \delta_{s,p,f,t} - 0.5\tau^2_{s,p,f} ) \\
R_{s,p,t}   &= \overline{R}_{s,p,t} \cdot \exp( \sigma_{s,p} \cdot \epsilon_{s,p,t} - 0.5\sigma^2_{s,p} ) 
\end{align}
where \(\bar{I}_{s,p,f,t}\) is the index without error defined
above, \(\tau_{s,p,f}\) is the log-normal observation error standard
deviation, \(\delta_{s,p,f,t}\) is the annual standard normal observation
error residual, \(\overline{R}_{s,p,t}\) is the equilibrium recruitment from
the Beverton-Holt stock-recruitment curve, \(\sigma_{s,p}\) is the recruitment
process error standard deviation, \(\epsilon_{s,p,t}\) is the annual standard
normal recruitment process error, and subscripts \(s,p,f,t\) are for species,
stock, fleet and year, respectively. Error is added to biomass indices
for pooled data independently of the error added to individual indices, i.e.,
\begin{align}
I_{s,f,t} &= \bar{I}_{s,f,t} \cdot \exp( \tau_{s,f} \cdot \delta_{s,f,t} - 0.5\tau^2_{s,f} ) \\
I_{p,f,t} &= \bar{I}_{p,f,t} \cdot \exp( \tau_{p,f} \cdot \delta_{p,f,t} - 0.5\tau^2_{p,f} ) \\
I_{f,t} &= \bar{I}_{f,t} \cdot \exp( \tau_{f} \cdot \delta_{f,t} - 0.5\tau^2_{f} ) 
\end{align}
where \(\tau_{s,f}, \tau_{p,f}, \tau_f\) were averaged over the components
of the pooled index.

\hypertarget{operating-model-data-quality-scenarios}{%
\subsubsection{Operating model data quality scenarios}\label{operating-model-data-quality-scenarios}}

The three data quality scenarios range from relatively data-rich to
data-poor by successively removing commercial CPUE index series from the
full set, i.e.,

\begin{enumerate}
\def\labelenumi{\arabic{enumi}.}
\tightlist
\item
  Data-\textbf{Rich}: Historical CPUE, Modern CPUE, Assemblage
  survey, Synoptic survey;
\item
  Data-\textbf{Mod}erate: Modern CPUE, Assemblage survey, Synoptic survey;
\item
  Data-\textbf{Poor}: Assemblage survey, Synoptic survey.
\end{enumerate}

To improve convergence, the Hierarchical Multi-stock and
Single-stock assessment models were initialised later under the \textbf{Mod}
and \textbf{Poor} data scenarios, with the starting year of the
assessments set to the first year with index data, which was 1984 in
HSHG for both scenario, and 1997 or 2003 for other areas under
the \textbf{Mod} and \textbf{Poor} scenarios, respectively.

\hypertarget{performance-evaluation}{%
\subsubsection{Performance evaluation}\label{performance-evaluation}}

\hypertarget{omniscient-manager-simulations}{%
\paragraph{Omniscient manager simulations}\label{omniscient-manager-simulations}}

Assessment model performance was measured against a simulated omniscient
fishery manager who is aware of all the future consequences of harvest
decisions and is, therefore, able to adapt the management to meet specific
quantitative objectives under any process error conditions
\citep{walters1998evaluation}. Omniscient manager solutions were used rather
than equilibrium based metrics \citep{punt2016management} because most stocks
were in a healthy state in 2016 (i.e.~above single-species \(B_{MSY}\),
Table 2) and, therefore, the time-path of fishery development was
important \citep{walters1988experimental}.

The omniscient manager was implemented as an optimisation
of future fishing effort by area (Appendix B), with the objective
function defined as
\begin{equation}
\mathcal{O}  = \left[\sum_{s,p}
                   -\log(\bar{C}_{s,p,\cdot}) \right] + 
                    \mathcal{P}_{diff}\left(\sum_p E_{p,\cdot}\right) +
                    \mathcal{P}_{init}\left(\sum_p E_{p,2017}\right), 
\end{equation}
where \(-\log \bar{C}_{s,p,\cdot}\) is the negative log of total
future catch for species \(p\) in area \(p\) over the projection period
(equivalent to maximising catch). Penalty functions \(\mathcal{P}\)
(eq. B.1) were applied for annual changes in total effort across all
three areas being above 20\% (\(\mathcal{P}_{diff}\)) to match the TAC
smoother in stochastic experiments, and differences greater than 10\%
between the last year of historical effort and the first year \(2017\) of
simulated effort (\(\mathcal{P}_{init}\)).

An omniscient manager solution was obtained for each stochastic
trajectory in the stochcastic management simulations. Each replicate was
run for 80 years to reduce end effects, such as transient dynamics at the
beginning of the projection, or a lack of consequences for overfishing
at the end of the projection.

\hypertarget{cumulative-catch-loss}{%
\paragraph{Cumulative catch loss}\label{cumulative-catch-loss}}

For each stochastic trajectory, the cumulative absolute loss
in catch was calculated as \citep{walters1998evaluation}:
\begin{align}
L_{s,p} &= \sum_{t = T_1}^{T_2} \vert C_{s,p,t,sim} - C_{s,p,t,omni} \vert,
\end{align}
where the \(C_{s,p,t,\cdot}\) values were commercial trawl catch for
species \(s\) and stock \(p\) from stochastic simulations (\(sim\)) or the
omniscient manager simulation (\(omni\)) simulation. When repeated over
all random seed values, the loss functions generated a distribution
of cumulative catch loss, which were then used to determine relative
performance of each assessment model under the three data scenarios.
Cumulative loss was calculated for the ten year period \(T_1 = 2028\) to
\(T_2 = 2037\), chosen in the middle of the projection period because
dynamics in the earlier time were dominated by the smoothers on effort
and catch for the omniscient manager and TACs, respectively, and after
2028 the omniscient manager's median effort has reached a stable
state near the multi-species optimum. Biomass loss was also
calculated, but inferences about assessment method performance in
preliminary simulations were not qualitatively different between
the two metrics, so catch loss was reported only, with biomass risk for
each stock indicated in reference to a critically overfished level,
defined as 40\% of their individual single-species \(B_{MSY}\) value
\citep{DFO2006A-Harvest-Strat}.

Cumulative absolute catch loss was used to calculate the
relative rank of each assessment model for each species/stock/OM scenario
combination, where lower loss ranked higher. Rank distributions
across species and stocks were then used to calculate an
average, minimum, and maximum rank of each assessment model to determine
the overall performance of each assessment model in each OM scenario.

\hypertarget{sensitivity-analyses}{%
\subsection{Sensitivity analyses}\label{sensitivity-analyses}}

Parameter prior distributions area a key feature of most contemporary
stock assessment models, even in data-rich contexts. Moreover, prior
distributions are a defining feature of hierachical multi-species
stock assessment models. Therefore, we focused sensitivity analyses on
fixed prior standard deviations for leading parameters \(B_{MSY}\) and \(U_{MSY}\), and the
hierarchical shrinkage prior SDs \(\tau_q\) and \(\sigma_{U_{MSY}}\) (Table 4).
Analyses were run with 50 simulation replicates for only data-rich and
data-poor scenarios.

\hypertarget{results}{%
\section{Results}\label{results}}

\hypertarget{omniscient-manager-performance}{%
\subsection{Omniscient Manager Performance}\label{omniscient-manager-performance}}

As expected, the omniscient manager was able to achieve the
theoretical multi-species optimal yield in the presence of
technical interactions (Figure 3, blue closed circle). Median
biomass, catch, and fishing mortality reach the equilibrium
after a transition period of about 20 years. During the
transitionary period, effort is slowly ramped up in each area
from the end of the historical period, stabilising around
area-specific \(E_{MSY}\) after about 12 years (Figure 4,
blue closed circles).

The multi-species optimal effort for the complex results in
overfishing Dover sole stocks between 53\% and 92\% of
area-/single-species \(B_{MSY}\) to increase fishery access to
English and Rock sole. Despite this tendency toward overfishing Dover
sole, very few optimal solutions risk severe overfishing of
Dover sole below 40\% of \(B_{MSY}\), indicating that lost Dover
sole yield from further overfishing relative to single-species
optimal levels is not compensated by increased English and
Rock sole yield (Table 5). The probability of being critically
overfished in the period 2028 - 2037 was 0\% for most stocks,
with two Dover sole stocks having 3\% (HSHG) and 1\% (QCS).

Although DER stocks begin the simulations in an overall healthy state,
the omniscient manager reduced fishing effort to near 0 in HSHG and QCS
areas early in the projection period in some replicates (Figure 4, HSHG
and QCS, 2016-2020). In these cases, anticipatory feedback control by
the omniscient manager reduced fishing effort to avoid low spawning stock
biomasses, thus and ensuring higher production in later time steps where
recruitment process error deviations were sustained at low levels.

\hypertarget{assessment-model-performance}{%
\subsection{Assessment model performance}\label{assessment-model-performance}}

\hypertarget{catch-loss}{%
\subsubsection{Catch loss}\label{catch-loss}}

Hierarchical Multi-stock assessment models ranked highest, on average, under
the Poor and Mod data quality scenarios (Table 6). Under the Poor data
quality scenario, the Hierarchical model had the lowest cumulative absolute
catch loss, on average, and ranked in the top three assessment methods
for all stocks. As data quantity increased for the Mod scenario, the average
absolute rank of the Hierarchical model degraded slightly, but still came
first in average rankings. It was only under the Rich scenario that the
Hierarchical model average rank fell to fourth place, although it still
ranked above the Single-stock model.

Under the Rich data quality scenario, the Species Pooling and Total
Aggregation methods ranked best, never ranking below fourth for
all stocks (Table 6, Rich). As data was removed for the Mod data
quality scenario, the average ranks of both Species Pooling and Total
Aggregation methods degraded, placing them second and third. The
Single-stock model had the worst average rank under all data-quality
scenarios, with its worst performance observed under the Mod data quality
scenario where it ranked 5th across all stocks.

Cumulative catch loss distributions for the Hierarchical model
and three data-pooling methods were fairly closely clustered
for the Rich and Mod data quality scenarios for most stocks (Figure 5). Only
the Single-stock model was qualitatively different, but this was not always
the case (e.g., HSHG Dover, Figure 5). Relative performance of each method
across data quality scenarios was stock dependent, with no consistent
trends across scenarios and stocks.

Technical interactions had both a positive and negative effect on meeting
the multi-species objectives and minimising catch loss. For
example, over the course of the projection period, the Hierarchical
Multi-stock model was able to bring median biomass levels close to the
multi-species optimal level \(B_{MSY,MS,s,p}\) for six out of the nine stocks
under the Poor data quality scenario (Figure 6, blue dots). The three
stocks that did not meet the multi-species optimal biomass were all in
the HSHG area, indicating that technical interactions may have been a
factor in the inability to meet the target harvest rates, choking off
TACs for some species in HSHG. For the HSHG area, catch of Dover and Rock
soles are choked by English sole TAC (Figure 7, HSHG, unfilled TAC bars),
which is low relative to the target level because the Hierarchical assessment
model viewed HSHG English sole as a larger and less productive stock than
it truly was, resulting in a persistent negative bias in the stock assessment
biomass. Furthermore, there is evidence that a large perturbation is needed
to improve the assessment for HSHG English sole and overcome the choke effect,
as the assessment model believes the TACs are appropriately scaled and biomass
is approaching \(B_{MSY,MS}\). For QCS and WCVI areas, all stocks were able
to meet the target harvest objectives and minimise loss directly because technical
interactions protected against overfishing. In QCS, English sole had close
to unbiased assessments, and was able to limit catches for Dover and Rock
soles despite large positive assessment errors for both (Figure 7, QCS).
Similarly, WCVI English sole catch was choked by TACs for Dover and Rock
sole, which both had unbiased assessments (Figure 7, WCVI).

\hypertarget{catch-biomass-trade-offs}{%
\subsubsection{Catch-Biomass trade-offs}\label{catch-biomass-trade-offs}}

Distributions of catch and biomass relative to \(MSY_{MS}\) and
\(B_{MSY,MS}\), respectively, were produced for the time
period \(2028 \leq t \leq 2037\) for each assessment model and Scenario
combination. The medians of those relative catch and biomass
distributions were visually compared to each other and to the
central 95\% of the omniscient manager's trajectories over the
same time period, to understand the biomass and catch trade-offs
between different model choices, and to compare each model to the
omniscient manager's optimal solution.

As indicated in the loss rankings, the Hierarchical Multi-stock
assessment model median biomass and catch between 2028 and 2037 came
closest (i.e., smaller Euclidean distance) to the omniscient manager median
levels under the Poor data quality scenario outside the HSHG area (Figure
8, compare points to the centre of the black crosshairs). Within HSHG,
the Hierarchical model tended to underfish relative to the omniscient
manager for the Poor data scenario, creating a large biomass surplus and
the lowest catch next to the Single-stock model, while the Spatial Pooling
method came closest to the omniscient manager for all three species.

Although the range of biomass-catch trade-offs are quite broad for each
stock, the majority of Scenario/Assessment combinations lie inside
the central 95\% distributions of the omniscient manager (Figure 8,
black crosshairs). Notable exceptions to this were (i) the Single-stock
model under the Rich data quality scenario, (ii) the Hierarchical model in
QCS under the Rich data quality scenario where Dover and Rock sole were
both critically overfished, (iii) the Spatial Pooling method in
WCVI under the Poor data quality scenario where Dover and English sole
were critically overfished, and (iv) the Total Aggregation and Species
Pooling methods under the Poor data quality scenario in QCS.

Catch-biomass trade-offs were approximately collinear under both
Rich and Poor data quality scenarios for all HSHG stocks, QCS English sole,
and WCVI Rock sole. For all these stocks except WCVI Rock sole,
spawning biomass is well above both the single-species and multi-species
optimal levels at the beginning of the projection period (e.g., Figure 6,
compare biomass at the start of the projection period to green open and
blue closed circles), meaning that the catch limits set by all methods
were depleting a standing stock and benefited from its surplus
production. Under these conditions, an increase in catch almost linearly
caused a decrease in biomass as the compensatory effect of density dependence
was minimal. A similar phenomenon explains the WCVI Rock sole collinearity,
but there the biomass was growing from a starting point above the
single-species \(B_{MSY,SS}\) towards the multi-species optimal level
\(B_{MSY,MS}\), asymptotically approaching it as biomass equilibrates to a
harvest rate well below the single species optimal rate \(U_{MSY,SS}\).

Catch-biomass trade-offs in the HSHG area indicated that Hierarchical
models performed more similar to the omniscient manager under the
Rich data quality scenario (Figure 8, HSHG). In the same area, the
Species Pooling method (ranked highest under the Rich scenario) produced
median biomass between 140\% and 210\% of the multi-species optimal level
(Figure 8, HSHS, Purple diamond), despite the superior performance with
respect to catch loss for Dover and English soles (indicated by proximity
to the horizontal median catch line).

As mentioned above, several stocks were pushed into a critically
overfished state by different methods. The Hierarchical Multi-stock
model (QCS, Rich data quality scenario) and Spatial Pooling method
(WCVI, Poor data quality scenario) produced median biomass levels
that were below 40\% of \(B_{MSY,SS}\) for two out of three species
in each area (Figure 8, QCS and WCVI). These methods avoided pushing
the QCS English and WCVI Rock soles, respectively, into
critically overfished states thanks largely to their relatively
low commercial catchability scalars. Low catchability means that their
multi-species optimal biomass \(B_{MSY,MS}\) was well above the
single-species optimal biomass level \(B_{MSY,SS}\), providing plenty of
room to overshoot the optimal biomass level while still avoiding the
critically overfished level. Even so, median biomass is still well
outside the omniscient manager's distribution for both stocks, with
catches far above the omniscient manager's median catch, ranging
between 150\% and 180\% of maximum yield under multi-species \(E_{MSY}\).
Finally, Total Aggregation and Spatial Pooling assessment methods had
higher than 10\% probability of being below 40\% of \(B_{MSY,SS}\) under the
Poor data scenario for (i) HSHG Dover sole (37 \% Total Aggregation,
42\% Species Pooling), (ii) QCS Dover sole (37 \% Total Aggregation,
11\% Species Pooling), (iii) HSHG English (12 \% Total Aggregation,
11 \% Species Pooling), and (iv) QCS Rock sole (11\% Total Aggregation).

\hypertarget{sensitivity-of-results-to-prior-standard-deviations}{%
\subsection{Sensitivity of results to prior standard deviations}\label{sensitivity-of-results-to-prior-standard-deviations}}

We summarised average model sensitivities by fitting linear regressions to
the distributions of median cumulative loss. To remove the effect of absolute
catch scales on the regressions, median loss distributions were standardised
across assessment models, stratified by species, stock-area, and data scenario.
Regressions with positive slopes have increasing catch loss with increasing prior
uncertainty, and negative or zero slopes indicate a decrease or no change in
catch loss with increasing prior uncertainty.

The Hierarchical Multi-stock model was least sensitive to increasing
\(B_{MSY}\) prior CVs in both Rich and Poor data quality scenarios.
Cumulative catch loss actually decreased on average with increasing
\(B_{MSY}\) prior CVs when catch limits were set by the Hierarchical
model under the Poor data quality scenario, decreasing by about 0.3
standard deviations over the range of CVs tested (Figure 9, left column,
Poor). While cumulative catch loss from Hierarchical models increased on
average with increasing \(B_{MSY}\) prior CVs under the Rich data quality
scenario, the slope of the regression line was the lowest among methods
(equal with the Species Pooling and Total Aggregation methods), implying
the least sensitivity to prior specifications.

Cumulative catch loss under the Hierarchical model was also
insensitive to the complex mean \(U_{MSY}\) hyperprior standard deviation,
with zero slope regression lines under both Rich and Poor data quality
scenarios over the range of tested log-normal prior SDs (Figure 9, middle
column). The lack of sensitivity of the Hierarchical model to the
\(U_{MSY}\) hyperprior was partially because target harvest rates were based
on theoretically optimal levels, and not estimated from the assessment model;
however, this implies that the biomass estimates were also insensitive to the
hyperprior standard deviation, which may be because the effect of the
hyperprior was soaked up by the two levels of shrinkage priors between
the complex mean and stock-specific estimates of \(U_{MSY}\).

Hierarchical Multi-stock assessment models were more able to suitably
scale catch limits to operating model biomass under relaxed catchability
and \(U_{MSY}\) hierarchical shrinkage priors. As \(\sigma_{U_{MSY}}\)
and \(\tau_q\) increased from 0.1 to 0.5, average catch loss dropped by about
0.4 standard deviations under the Rich data quality scenario, and 0.8 standard
deviations for the Poor data quality scenario (Figure 9, third column).
Improvements stemmed from a change in assessment model errors in the HSHG area,
in particular (Figures C.1, C.2, Appendix C). The change in assessment
performance switched the choke species from English to Rock sole in HSHG,
and while most assessments were biased and had strong retrospective
patterns, the combination of higher catch limits and favourable
technical interactions produced catches that were more closely aligned
with the omniscient manager's and the multi-species maximum yield.

As expected, cumulative catch loss increased with decreasing prior
precision under all other assessment methods. The Single-stock model
was most sensitive to \(B_{MSY}\) CVs under both Rich and Poor data quality
scenarios, and to \(U_{MSY}\) log-normal SDs under the Poor data quality scenario,
indicating that the single-stock model required the most prior knowledge
for scaling biomass estimates correctly. The Total Aggregation and Spatial
Pooling methods were most sensitive to the \(U_{MSY}\) prior standard deviation,
which may indicate a that the aggregate prior mean values we calculated for
those assessment models were not well supported by the aggregated data.

\hypertarget{discussion}{%
\section{Discussion}\label{discussion}}

In this paper, we demonstrated that hierarchical stock assessment
models may improve management performance in a data-limited, multi-species
flatfish fishery. When available data quality was moderate or poor
(indicated here by time-series length), biomass estimates from
hierarchical stock assessment models resulted in catches that were closer
to an omniscient manager's optimal reference
series compared to catch limits derived from single-stock and data-pooling
assessment methods. Under high data quality scenarios, data-pooling
methods outperformed hierarchical models, but the latter
still outperformed single-stock assessment methods. Improved performance
relative to single-stock models under lower data quality conditions
is consistent with our previous study, where statistical performance
of hierarchical multi-stock assessments improved with decreasing
data quantity and quality \citep{johnson2018evaluating}. This suggests
that hierarchical assessment methods would generally be a better
approach than conventional single-species methods under typical
fisheries data quality conditions.

Our results arise from models that are necessarily a simplification
of the real stock-management system. The harvest rules applied to DER complex
species were relatively simple and may require more detail or complexity
for practical applications. First, the harvest rules were all constant target
harvest rates, which do not include precautionary ``ramping-down'' of
catch towards a limit biomass level \citep{DFO2006A-Harvest-Strat, cox2013roles}.
Including a ramped harvest rule may have reduced the probability of
some stocks being critically overfished in some cases, but probably at
some further cost of choke effects. Second, catch
limits for the simulated DER complex were set based on fixed target harvest
rates that were derived \emph{a priori} from multi-species yield curves, and not
estimated as part of the assessment models that we tested. Incorporating
multi-species yield curve calculations based on assessment model ouput
into the harvest decision would be simple to do, but would require either
a model of increased complexity to link fishing effort to single-species
yield, or an extra assumption linking effort to surplus production model
yield calculations, which would likely increase assessment model errors.
Finally, the TAC allocation model for data-pooled methods was only one
example from a large set of potential options. Understanding the relative
risks of data-pooling would require testing alternative allocation
methods, which was beyond the scope of this paper.

We only considered multi-species technical interactions, which although
an important part of exploited system dynamics, are not the entire
story. Although there is limited evidence for ecological interactions among
DER complex species \citep{pikitch1987use, wakefield1984feeding}, what does exist
may influence the multi-species yield relationship with fishing effort or, as
with technical interactions, inhibit the ability of the management system to
meet target catch levels. For example, individual survival or growth may change
in response to varied fishing pressure due to unmodeled linkages
\citep{collie2001biological}. Yet, including such ecological interactions would
imply a highly data rich scenario, which is counter to our focus on
data-limited, multi-species fisheries. Furthermore, accounting for potential
ecological interactions would require multiple OMs to test performance against
a range of plausible hypotheses since ecological uncertainties are much
broader in complexity and scope than technical interactions alone. Nevertheless,
future work combining technical interactions with minimum realistic models
for ecological interactions could help determine the extent to which assessment
approaches affect these more complex multi-species fisheries outcomes
\citep{punt1995effects}. For example, while diet overlap between the three
species is small off the coast of Oregon, the major Rock sole prey
was recently settled pleuronectiform fishes, which may include Dover and
English sole young and therefore shift the complex equilibrium as fishing
pressure is applied, reducing predation mortality for Dover and English
sole and prey availability for Rock sole \citep{wakefield1984feeding, collie2001biological}.

Our effort model applied to the DER complex was also a simplification of
reality, where effort was limited only by the TACs in each area.
Limiting by TACs was intended to reflect the management of the real BC
groundfish fishery in which harvester decisions drive TAC utilisation
among target species \citep[via increasing catchability;][]{punt2011calculating},
and non-target or choke species \citep[via decreasing catchability;][]{branch2008matching}. Changing catchability for targeting or avoidance
could be simulated as a random walk in the projections,
with correlation and variance based on the historical period, or perhaps
simulated via some economic sub-model that accounted for ex-vessel
prices and variable fishing costs. These economic factors could affect
targeting and avoidance behaviour among species \citep{punt2011calculating, punt2020multispecies}, as well as effort allocation among stock-areas
\citep{hilborn1987general, walters1999multispecies}; however, it is not
clear that our median results would be significantly different given
the potential magnitude of assessment model errors in data-limited
scenarios. Impacts of a detailed effort dynamics sub-model would probably
be more important in more extreme data-limited scenarios that relied
solely on fishery CPUE as an index of abundance. In fact, it would be
interesting to determine whether the hierarchical information-sharing
approach would exacerbate assessment model errors in such a (common)
context where fishery CPUE is the main abundance index.

Despite the limitations above, our results indicate that even in fisheries
with long time series of catch and effort data, hierarchical multi-species
assessment models may be preferable over typical single-species methods.
The poor performance of the single-species models in all scenarios
highlights the difference between data-rich (i.e., a higher quantity of data)
and information-rich (i.e., data with higher statistical power) fisheries.
The data-rich scenario differed
from data-moderate and data-poor scenarios by the inclusion of a historical
series of fishery dependent CPUE, which was quite noisy and subject to the
effects of changing harvester behaviour like targeting (variable catchability),
and therefore, additional historical CPUE data had little effect on cumulative
catch loss under the single-species models. In contrast, the data-pooling
procedures all ranked higher than single-species and multi-species models
under the data-rich scenario, as they were able to leverage additional
statistical power from the historical CPUE by effectively increasing the
sample size through data aggregation. The superior performance of the
hierarchical model over the single-species model under the data-rich scenario
indicate that shared priors partially compensate for low statistical power,
but not as much as data-pooling.

Although the data-pooled methods performed better under the rich scenario,
they were more sensitive to priors, so those results may be optimistically
biased in all scenarios. Data-pooled observation errors were simulated as
independent of the observation errors in the component indices, using the
average standard deviation of the components. If aggregate indices pooled
errors from each component index, then the resulting observation error
variance would be additive in the components, especially if those errors
were positively correlated, which may be the case under a common survey
or fishery.

A dual effect of control was observed under catch limits set by
the hierarchical models in the data-poor scenario, in which lower
contrast in assessment model outputs reduced the statistical power of
assessment model data, sacrificing long term adaptability of the management
system in favour of short term stability in catch. Moreover, there was
limited evidence that adaptive learning was facilitated through the shared
hierarchical prior distributions \citep{collie1991adaptive}. Uninformative catch
series resulting from a lack of large perturbations (sometimes caused by
large catch errors) resulted in hierarchical multi-species assessment model
estimates that viewed HSHG English sole as a larger and less productive
stock, causing negatively biased assessment estimates of biomass to approach
the multi-species \(B_{MSY,MS}\) value for that stock, and indicating that
the assessments believed the TACs were appropriately scaled to the target
harvest rates. Relaxing hierarchical prior standard deviations in the
sensitivity analysis removed this persistent bias and all stocks approached
biomass levels associated with multi-species maximum yield; however,
information sharing was not wholly responsible for the improved behaviour,
which appeared to also rely on a favourable combination of assessment errors
and technical interactions as in the other stock areas. Adaptive learning catalysts
like large stock perturbations may have been limited by a low recruitment
process error standard deviation of \(\sigma_{s,p} = .4\) in DER complex
simulations, which was required for fitting the DER complex OM to data.
Future work could relax this assumption to test whether perturbations
stemming from higher recruitment variability would improve adaptive
learning of hierarchical assessment methods \citep{walters1986adaptive}.

We showed that choke effects are not a uniformly negative outcome for
multi-species fisheries, and may indicate a mismatch between the target
harvest rate and optimal complex yield. The usual assumption is that
choke species restrict access to fishing grounds, decreasing profitability
through lost yield of target species, and higher quota prices for choke species
\citep{mortensen2018identifying}; however, we found that choke species
sometimes prevented overfishing when TACs for the non-choke species were
set too high, allowing harvest strategies to meet multi-species
objectives despite large assessment errors for individual species
in the complex. In reality, choke effects would likely be lessened
by changing species catchability via harvester targeting and avoidance,
creating a more complex relationship between effort and complex
yield; but, the existence of a choke species would still indicate
a mismatch between an individual species TAC and the optimal exploitation
level for the multi-species complex.

\hypertarget{conclusion}{%
\subsection{Conclusion}\label{conclusion}}

Hierarchical multi-species assessment models can out-perform single-species
assessment models in meeting multi-species harvest objectives across
data-rich, data-moderate, and data-poor scenarios. As expected, biomass
estimation performance of hierarchical models improved relative to
other methods as data quantity was reduced, and - as hoped - this
translated into improved management performance across the multi-species
flatfish fishery. We therefore recommend that assessment and management
of multi-species fisheries include technical interactions when
designing harvest strategies and management procedures aimed at
achieving strategic objectives. Otherwise, error-prone single-species
approaches may give a misleading picture of the expected performance
of multi-species fishery management.

\hypertarget{acknowledgements}{%
\section{Acknowledgements}\label{acknowledgements}}

Funding for this research was provided by a Mitacs Cluster Grant to S.P. Cox
in collaboration with the Canadian Groundfish Research and Conservation
Society, Wild Canadian Sablefish, and the Pacific Halibut Management
Association. We thank S. Anderson and M. Surry at the Fisheries and Oceans,
Canada Pacific Biological Station for fulfilling data requests. Further
support for S.P.C. and S.D.N.J. wass provided by an NSERC Discovery Grant
to S.P. Cox.

\clearpage

\hypertarget{tables}{%
\section{Tables}\label{tables}}

\begin{table}[!h]

\caption{\label{tab:dataSources}Fishery dependent and independent indices of 
biomass available for stock assessments of DER complex species.}
\centering
\resizebox{\linewidth}{!}{
\begin{tabular}[t]{lll}
\toprule
Series & Extent & Description\\
\midrule
Historical Commercial & 1976 - 1995 & Historical period commercial CPUE\\
Modern Commercial & 1996 - 2016 & Modern period commercial CPUE\\
HS Assemblage & 1984 - 2002 & Hecate Strait Assemblage trawl survey biomass index, biennial\\
Synoptic & 2003 - 2016 & Multi-species Synoptic trawl survey biomass index, biennial\\
\bottomrule
\end{tabular}}
\end{table}

\clearpage

\begin{table}[!h]

\caption{\label{tab:stockStatusTab}Unfished biomas $B_0$, single-species MSY based 
reference points $B_{MSY,SS}$ and $U_{MSY,SS}$, stock status as absolute
biomass in 2016 $B_{2016}$, depletion relative to single-species
optimal biomass $B_{2016}/B_{MSY,SS}$, and commercial trawl 
catchability scalar $q^{F}$ for all DER complex stocks in 2016. Biomass
quantities are given in kilotonnes, and depletion and harvest rates
are unitless.}
\centering
\resizebox{\linewidth}{!}{
\begin{tabular}[t]{ccccccc}
\toprule
\multicolumn{1}{c}{ } & \multicolumn{3}{c}{Single-species Reference Points} & \multicolumn{2}{c}{Stock Status} & \multicolumn{1}{c}{Comm. Catchability} \\
\cmidrule(l{3pt}r{3pt}){2-4} \cmidrule(l{3pt}r{3pt}){5-6} \cmidrule(l{3pt}r{3pt}){7-7}
Stock & $B_0$ & $B_{MSY,SS}$ & $U_{MSY,SS}$ & $B_{2016}$ & $B_{2016}/B_{MSY,SS}$ & $q^{F}$\\
\midrule
\addlinespace[0.3em]
\multicolumn{7}{l}{\textbf{Dover sole}}\\
\hspace{1em}HSHG & 17.50 & 6.06 & 0.17 & 6.06 & 1.00 & 0.022\\
\hspace{1em}QCS & 5.98 & 1.99 & 0.15 & 1.74 & 0.88 & 0.017\\
\hspace{1em}WCVI & 15.20 & 5.03 & 0.19 & 4.05 & 0.81 & 0.039\\
\addlinespace[0.3em]
\multicolumn{7}{l}{\textbf{English sole}}\\
\hspace{1em}HSHG & 9.98 & 3.42 & 0.31 & 5.95 & 1.74 & 0.024\\
\hspace{1em}QCS & 0.57 & 0.19 & 0.29 & 0.38 & 2.04 & 0.011\\
\hspace{1em}WCVI & 0.90 & 0.30 & 0.29 & 0.38 & 1.26 & 0.042\\
\addlinespace[0.3em]
\multicolumn{7}{l}{\textbf{Rock sole}}\\
\hspace{1em}HSHG & 16.35 & 5.75 & 0.23 & 8.40 & 1.46 & 0.009\\
\hspace{1em}QCS & 5.57 & 1.90 & 0.21 & 1.63 & 0.86 & 0.014\\
\hspace{1em}WCVI & 1.72 & 0.63 & 0.23 & 0.68 & 1.09 & 0.009\\
\bottomrule
\end{tabular}}
\end{table}

\clearpage

\begin{table}[!h]

\caption{\label{tab:AMpriorTab}Prior distributions used for each assessment
model. For single stock and hierarchical multi-stock assesssment
models, $s_f$ is 1 for commercial indices, and 0.5 for survey
indices.}
\centering
\begin{tabular}[t]{l}
\toprule
Prior\\
\midrule
\addlinespace[0.3em]
\multicolumn{1}{l}{\textbf{Totally aggregated AM}}\\
\hspace{1em}$B_{MSY} \sim N(\mu_{B_{MSY}}, 0.2 \cdot \mu_{B_{MSY}})$\\
\hspace{1em}$\log U_{MSY} \sim N(\mu_{\log U_{MSY}}, 0.4)$\\
\hspace{1em}$\log q_{f} \sim N(\mu_{\log q_{f}}, 1.0)$\\
\addlinespace[0.3em]
\multicolumn{1}{l}{\textbf{Species-Pooled AM}}\\
\hspace{1em}$B_{MSY,p} \sim N(\mu_{B_{MSY,p}}, 0.2 \cdot \mu_{B_{MSY,p}})$\\
\hspace{1em}$\log U_{MSY,p} \sim N(\mu_{\log U_{MSY,p}}, 0.4)$\\
\hspace{1em}$\log q_{p,f} \sim N(\mu_{\log q_{p,f}}, 1.0)$\\
\addlinespace[0.3em]
\multicolumn{1}{l}{\textbf{Spatial-Pooled AM}}\\
\hspace{1em}$B_{MSY,s} \sim N(\mu_{B_{MSY,s}}, 0.2 \cdot \mu_{B_{MSY,s}})$\\
\hspace{1em}$\log U_{MSY,s} \sim N(\mu_{\log U_{MSY,s}}, 0.4)$\\
\hspace{1em}$\log q_{s,f} \sim N(\mu_{\log q_{s,f}}, 1.0)$\\
\addlinespace[0.3em]
\multicolumn{1}{l}{\textbf{Single-stock AM}}\\
\hspace{1em}$B_{MSY,s,p} \sim N(\mu_{B_{MSY}}, 0.1 \cdot \mu_{B_{MSY}})$\\
\hspace{1em}$\log U_{MSY,s,p} \sim N(\mu_{\log U_{MSY}}, 0.1)$\\
\hspace{1em}$\log q_{s,p,f} \sim N(\mu_{\log q_{s,p,f}}, s_f)$\\
\addlinespace[0.3em]
\multicolumn{1}{l}{\textbf{Hierarchical multi-stock AM}}\\
\hspace{1em}$B_{MSY,s,p} \sim N(\mu_{B_{MSY}}, 0.2 \cdot \mu_{B_{MSY}})$\\
\hspace{1em}$\log U_{MSY,s,p} \sim N(\log \overline{U}_{MSY,s}, \sigma_{U_{MSY}})$\\
\hspace{1em}$\log \overline{U}_{MSY,s} \sim N(\log \overline{U}_{MSY}, \sigma_{U_{MSY}})$\\
\hspace{1em}$\log \overline{U}_{MSY} \sim N( \mu_{\log U_{MSY}}, 0.4)$\\
\hspace{1em}$\log q_{s,p,f} \sim N(\log \overline{q}_{s,f}, \tau_q)$\\
\hspace{1em}$\log \overline{q}_{s,f} \sim N( \mu_{\log q_{s,f}}, s_f)$\\
\addlinespace[0.3em]
\multicolumn{1}{l}{\textbf{All AMs}}\\
\hspace{1em}$\zeta_{s,p,t} \sim N(0,0.05)$\\
\bottomrule
\end{tabular}
\end{table}

\clearpage{}

\begin{table}[!h]

\caption{\label{tab:sensRunsTable}Summary of sensitivity analyses, showing
the total number of experiments, the factor being varied, the
levels of that factor, and the data scenarios and AMs included
in the analysis.}
\centering
\resizebox{\linewidth}{!}{
\begin{tabular}[t]{lllll}
\toprule
N & Factor & Levels & Scenarios & AMs\\
\midrule
30 & $B_{MSY}$ prior CV & $0.1, 0.5, 1.0$ & Rich, Poor & All\\
30 & $U_{MSY}$ prior SD & $0.1, 0.5, 1.0$ & Rich, Poor & All\\
6 & $\tau_q$, $\sigma_{U_{MSY}}$ & $0.1, 0.2, .5$ & Rich, Poor & Hierarchical only\\
\bottomrule
\end{tabular}}
\end{table}

\begin{landscape}

\begin{table}[!h]

\caption{\label{tab:omniPerfTab}Probability of being overfished and experiencing
overfishing with respect to single-species reference points, and 
catching less than the historical miniumum during the time period
2028 - 2037 for all nine DER complex stocks when managed by the 
omniscient manager.}
\centering
\resizebox{\linewidth}{!}{
\begin{tabular}[t]{cccccc}
\toprule
\multicolumn{1}{c}{ } & \multicolumn{2}{c}{Prob. of being overfished} & \multicolumn{2}{c}{Prob. of overfishing} & \multicolumn{1}{c}{Prob. of low catch} \\
\cmidrule(l{3pt}r{3pt}){2-3} \cmidrule(l{3pt}r{3pt}){4-5} \cmidrule(l{3pt}r{3pt}){6-6}
Stock & $P(B_t < .4B_{MSY,SS})$ & $P(B_t < .8B_{MSY,SS})$ & $P(C_t > MSY_{SS})$ & $P(F_t > F_{MSY,SS})$ & $P(C_t < \min C_{1951:2016})$\\
\midrule
\addlinespace[0.3em]
\multicolumn{6}{l}{\textbf{Dover sole}}\\
\hspace{1em}HSHG & 0.03 & 0.70 & 0.74 & 0.97 & 0\\
\hspace{1em}QCS & 0.01 & 0.46 & 0.66 & 0.86 & 0\\
\hspace{1em}WCVI & 0.00 & 0.23 & 0.54 & 0.63 & 0\\
\addlinespace[0.3em]
\multicolumn{6}{l}{\textbf{English sole}}\\
\hspace{1em}HSHG & 0.00 & 0.12 & 0.64 & 0.49 & 0\\
\hspace{1em}QCS & 0.00 & 0.00 & 0.24 & 0.00 & 0\\
\hspace{1em}WCVI & 0.00 & 0.03 & 0.53 & 0.23 & 0\\
\addlinespace[0.3em]
\multicolumn{6}{l}{\textbf{Rock sole}}\\
\hspace{1em}HSHG & 0.00 & 0.01 & 0.50 & 0.06 & 0\\
\hspace{1em}QCS & 0.00 & 0.04 & 0.53 & 0.29 & 0\\
\hspace{1em}WCVI & 0.00 & 0.00 & 0.01 & 0.00 & 0\\
\bottomrule
\end{tabular}}
\end{table}

\end{landscape}

\clearpage

\begin{table}[!h]

\caption{\label{tab:lossRankTable}Summary of AM rankings with respect to cumulative
absolute catch loss between 2026 and 2035 under each scenario. The
rank column shows the average rank of the AMs over all species/areas,
and the range of that distribution in parentheses. AMs are ordered by
mean rank within a scenario.}
\centering
\begin{tabular}[t]{cc}
\toprule
AM & Average Loss Rank (range)\\
\midrule
\addlinespace[0.3em]
\multicolumn{2}{l}{\textbf{Rich}}\\
\hspace{1em}Species Pooling & 1.67 (1, 3)\\
\hspace{1em}Total Aggregation & 2.33 (1, 4)\\
\hspace{1em}Spatial Pooling & 3.00 (2, 4)\\
\hspace{1em}Hierarchical Multi-stock & 3.22 (1, 5)\\
\hspace{1em}Single Stock & 4.78 (4, 5)\\
\addlinespace[0.3em]
\multicolumn{2}{l}{\textbf{Mod}}\\
\hspace{1em}Hierarchical Multi-stock & 2.00 (1, 4)\\
\hspace{1em}Species Pooling & 2.11 (1, 4)\\
\hspace{1em}Total Aggregation & 2.44 (1, 4)\\
\hspace{1em}Spatial Pooling & 3.44 (2, 4)\\
\hspace{1em}Single Stock & 5.00 (5, 5)\\
\addlinespace[0.3em]
\multicolumn{2}{l}{\textbf{Poor}}\\
\hspace{1em}Hierarchical Multi-stock & 1.44 (1, 3)\\
\hspace{1em}Spatial Pooling & 2.67 (1, 5)\\
\hspace{1em}Total Aggregation & 3.33 (2, 5)\\
\hspace{1em}Single Stock & 3.67 (1, 5)\\
\hspace{1em}Species Pooling & 3.89 (2, 5)\\
\bottomrule
\end{tabular}
\end{table}

\clearpage

\hypertarget{figures}{%
\section{Figures}\label{figures}}

\begin{figure}[htb]

{\centering \includegraphics[width=6in]{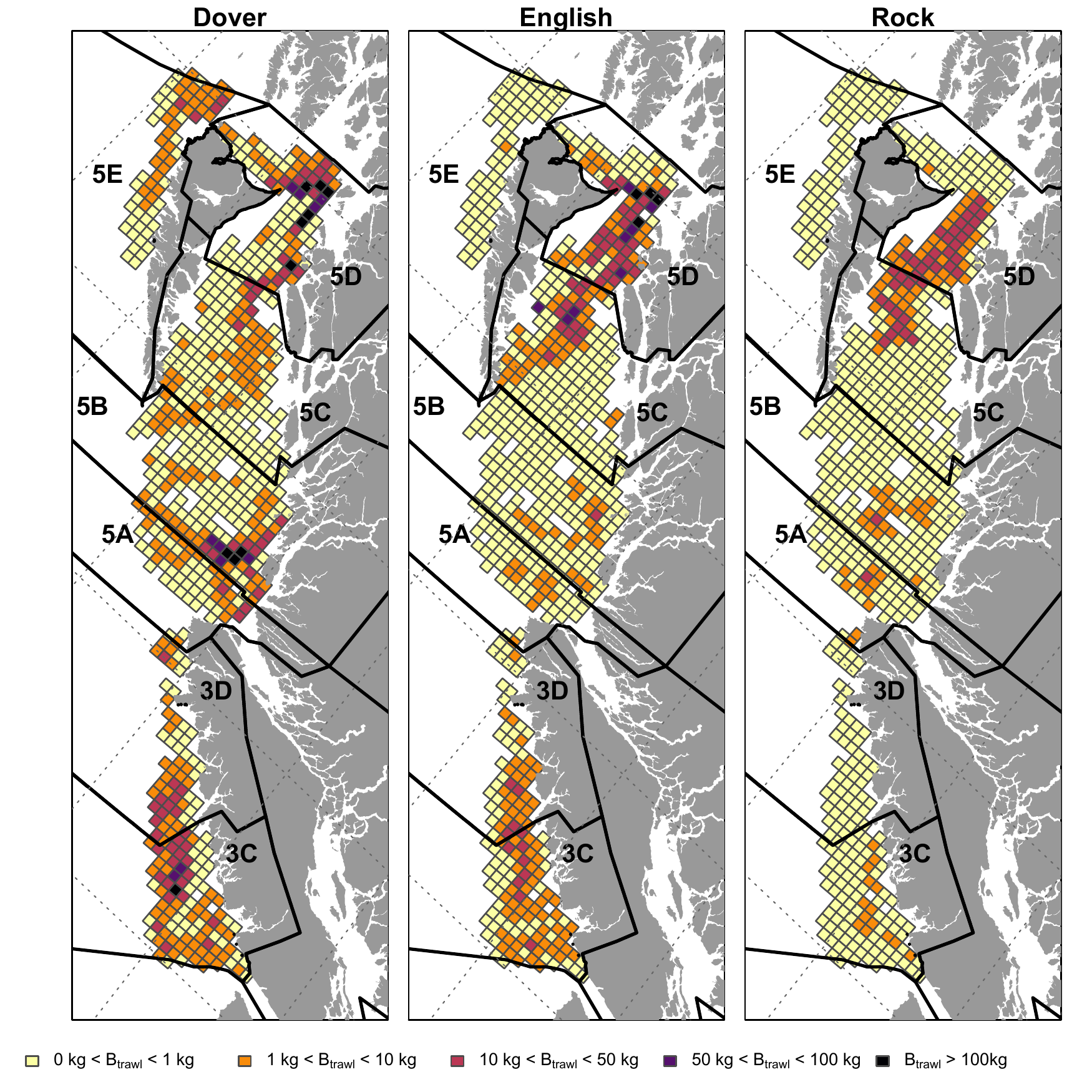} 

}

\caption{Mininum trawlable survey biomass $B_{trawl}$ estimates for DER 
complex species on the BC coast, aggregated to a 10km square grid.
Estimates are produced by scaling average trawl survey ($kg/m^2$) 
density values in each grid cell by the cell's area in $m^2$. 
Locations that do not show a coloured grid cell do not have any 
survey blocks from which to calculate relative biomass. Survey density 
for each grid cell is calculated from data for the Hecate Strait 
Assemblage Survey and the BC Groundfish Trawl Synoptic Survey, stored 
in the GFBio data base maintained at the Pacific Biological Station 
of Fisheries and Oceans, Canada. Thick black lines delineate the major 
statistical areas 3CD and 5ABCDE used for groundfish management BC, while
the dashed grey lines makr out latitude and longitude, indicating the
rotation of the coordinates to save space. The full colour figure 
is available in the online version of the article.}\label{fig:fig1-DERccomplexSurveyBio}
\end{figure}

\clearpage

\begin{figure}[htb]

{\centering \includegraphics[width=6in]{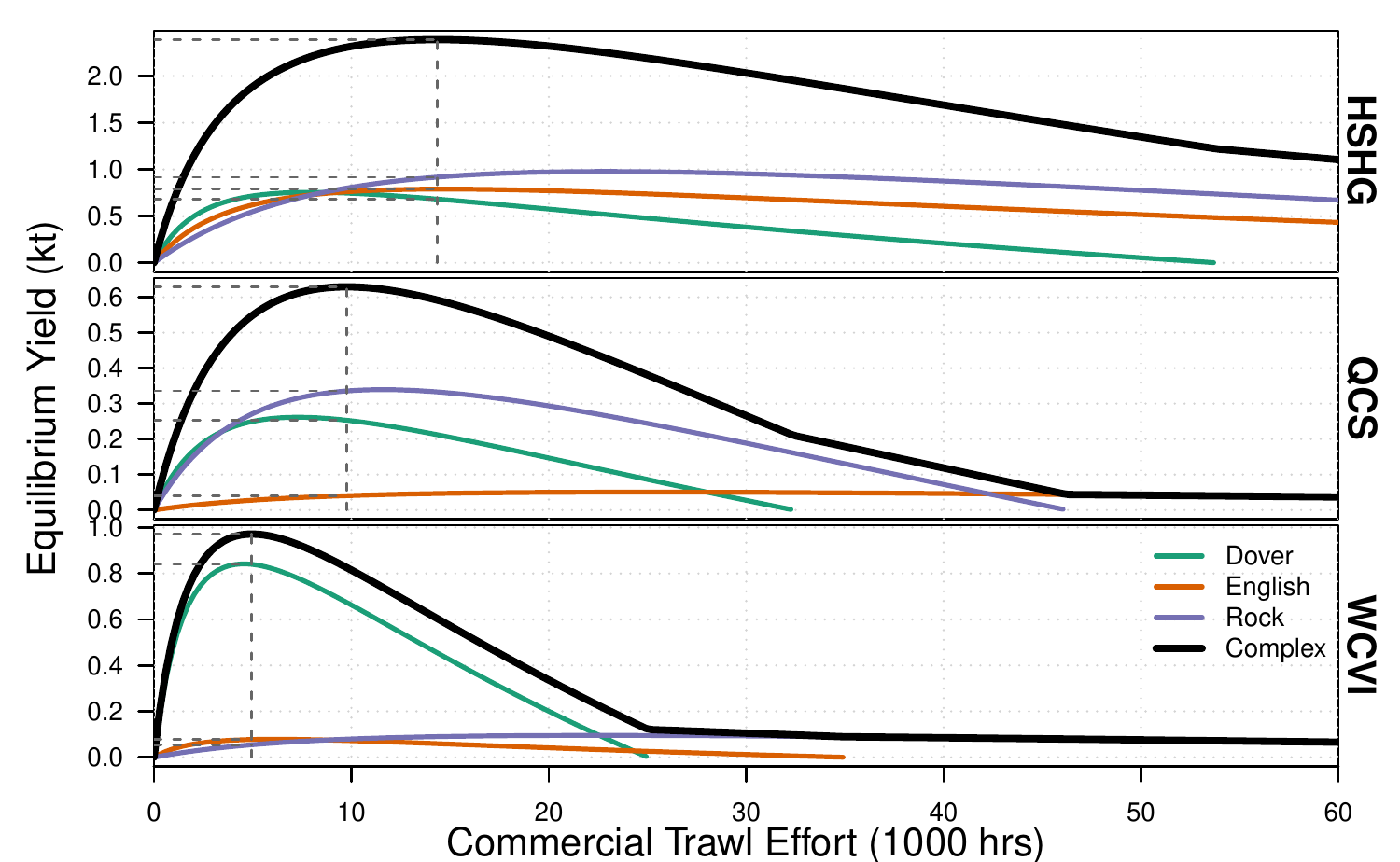} 

}

\caption{Operating model equilibrium yield curves for a given
fishing effort for the nine DER complex management units, as well as 
complex yield curves within each stock area. Each panel shows the three
individual DER complex species yield curves within a given stock area in
different colours, and the complex yield curve found by summing the three
yield curves as a thick black line.}\label{fig:fig2-complexYieldCurves}
\end{figure}

\clearpage

\begin{landscape}{}

\begin{figure}[htb]

{\centering \includegraphics[width=9in]{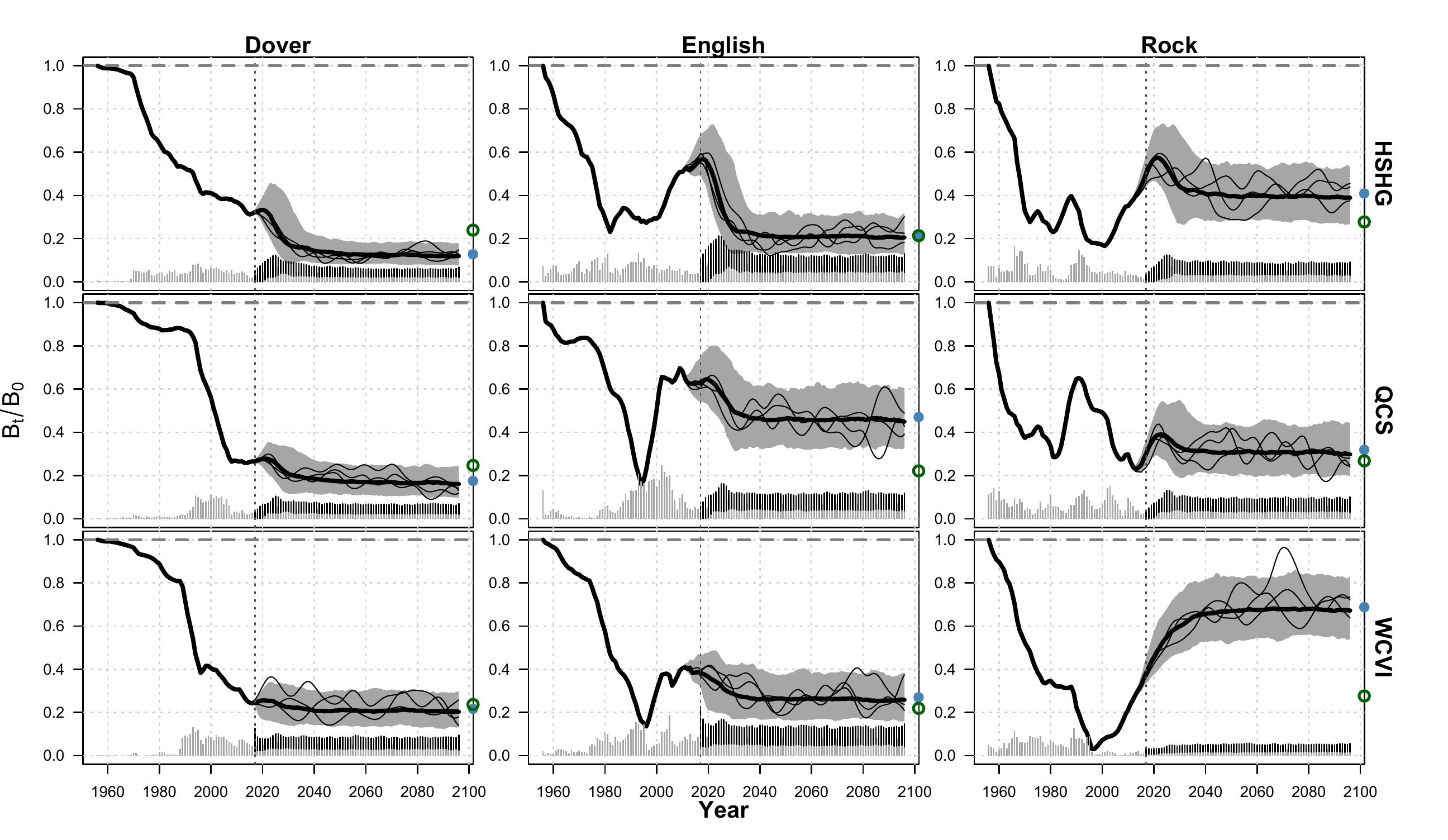} 

}

\caption{Spawning biomass depletion and relative catch 
simulation envelopes for all nine DER complex management units from 
the omniscient manager simulations. Median biomass is shown by the thick 
black line, with the grey region showing the central 95\% of the 
distribution of spawning biomass, and thin black lines 
showing three randomly selected simulation replicates. Catch is shown as grey 
bars in the historical period, which represent median catch in the projection, 
with thin vertical line segments showing the central 95\% of the catch 
distribution. The depletion level associated with the traditional single 
species optimal biomass $B_{MSY,SS}$ is shown as a dashed horizontal green 
line, while the depletion level associated with our derived complex 
level multi-species optimal yield is shown as a blue horizontal dashed line.
}\label{fig:fig3-omniBtCtTulips}
\end{figure}

\clearpage

\end{landscape}

\begin{figure}[htb]

{\centering \includegraphics[width=6in]{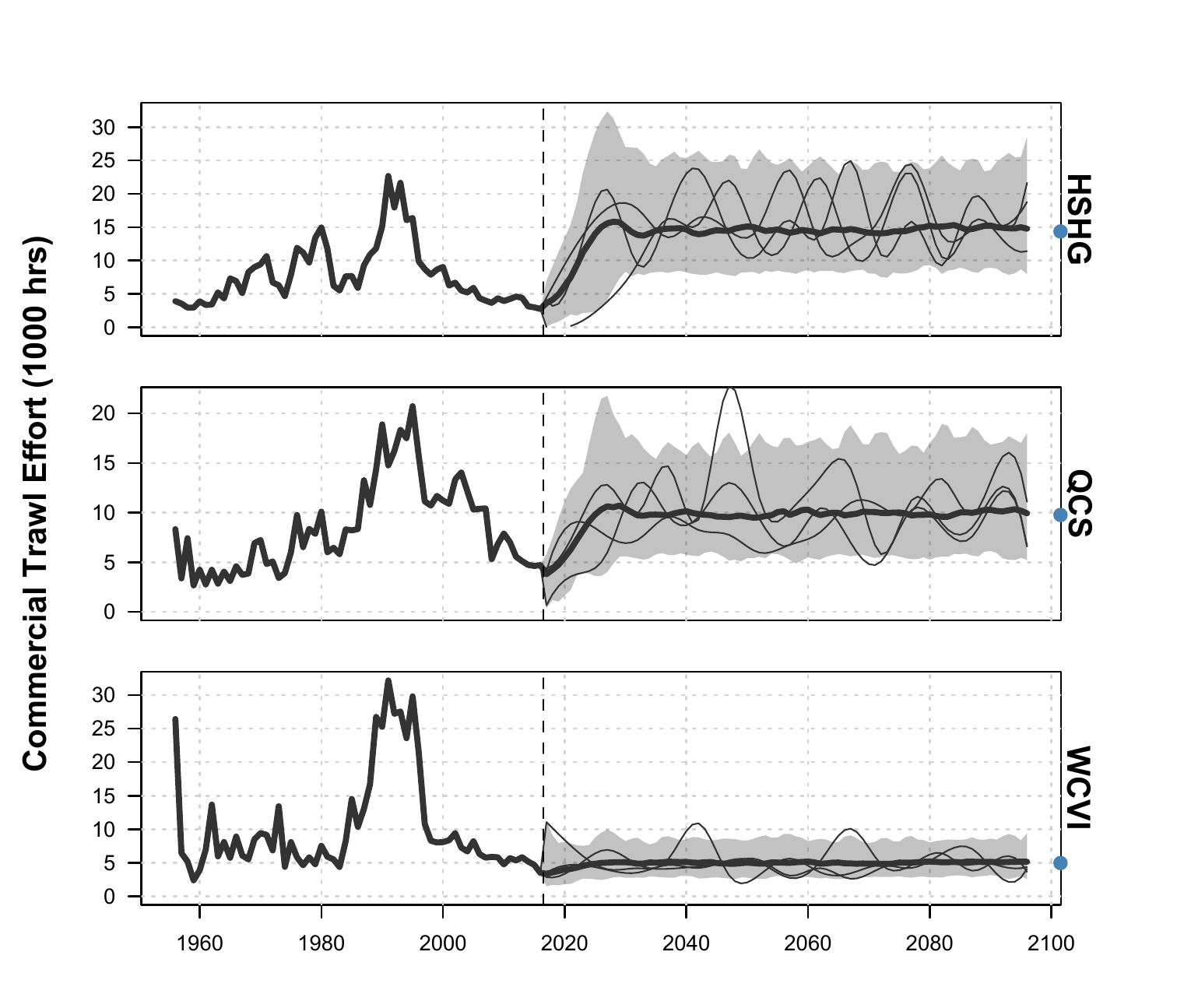} 

}

\caption{Commercial fishing effort simulation envelopes for 
each stock area. Historical and median simulated effort in the projection 
period are shown by a thicker black line, while the central 95\% of the 
distribution of simulated effort is shown as grey shaded region in the 
projection period. Single simulation replicates in the projection period 
are shown as thinner black lines.}\label{fig:fig4-omniEffTulips}
\end{figure}

\begin{landscape}

\clearpage

\begin{figure}[htb]

{\centering \includegraphics[width=9in]{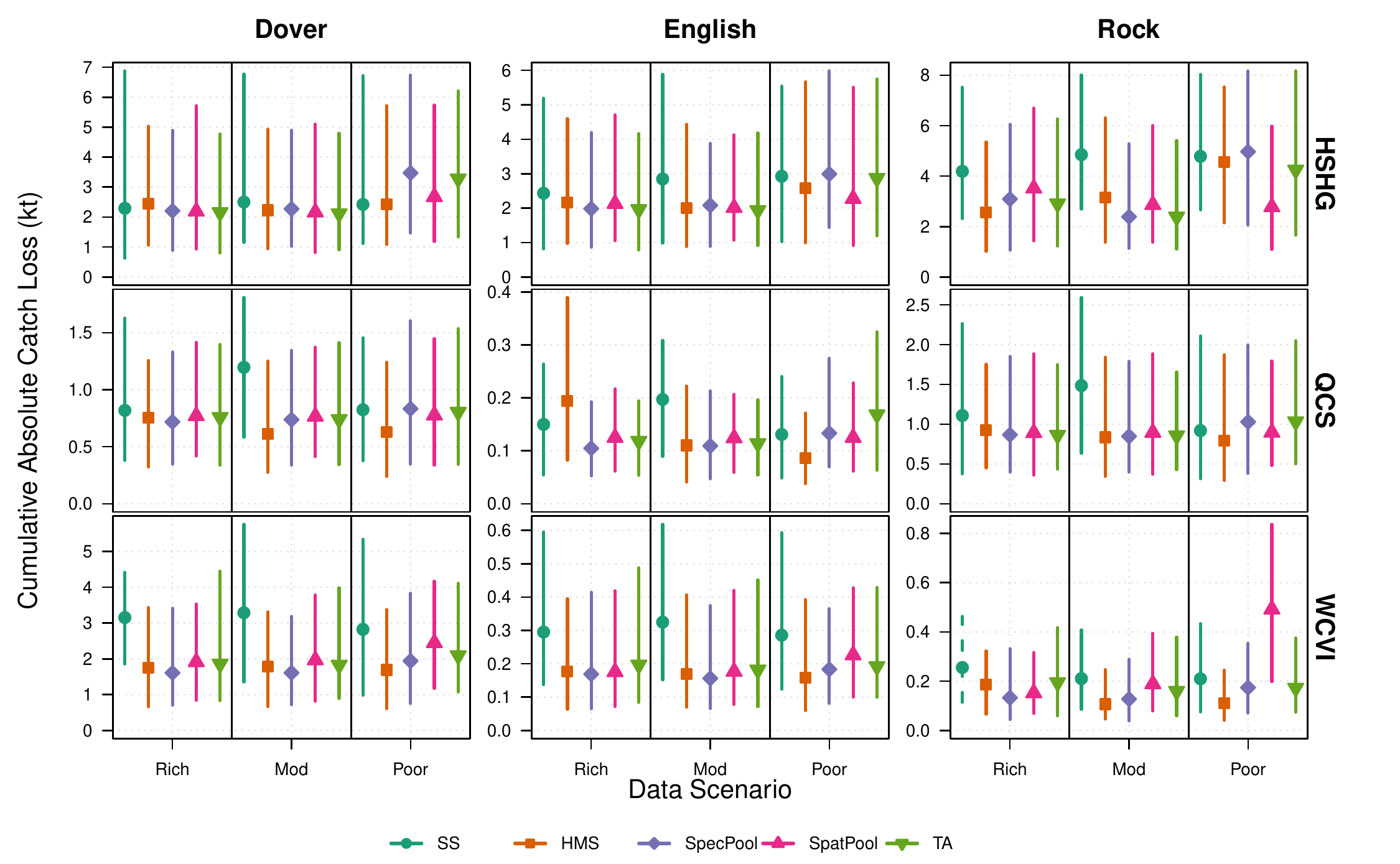} 

}

\caption{Distributions of cumulative absolute loss in catch (kt) 
for the projection years 2026 to 2036 under each assessment model and
OM data scenario (x-axis labels). The points show the median loss, 
while line segments show the central 90\% of cumulative loss 
distributions. Panels show the cumulative loss distributions by 
species (columns) and stocks (rows). Each assessment model point type and 
colour is shown in the legend at the bottom. Loss distributions
for assessment models that couldn't reach 100 replicates with satisfactory
convergence are shown as a dashed line.}\label{fig:fig5-batchLossAbsCatch}
\end{figure}

\clearpage

\begin{figure}[htb]

{\centering \includegraphics[width=9in]{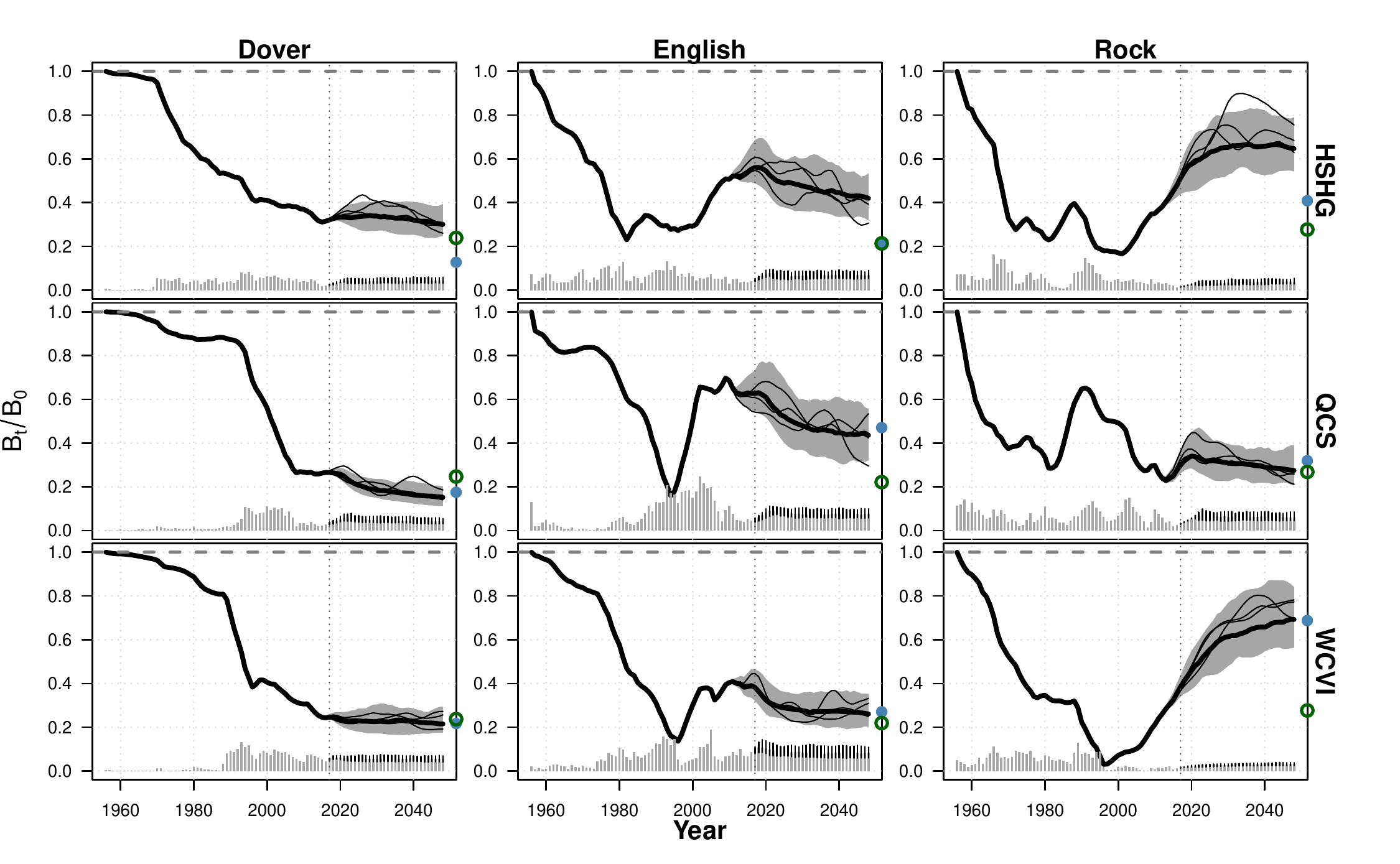} 

}

\caption{Spawning biomass depletion and relative catch 
simulation envelopes for all nine DER complex management units when 
assessed by the Hierarchical Multi-stock assessment model under the 
Poor data-quality scenario. Median biomass is shown by the thick 
black line, with the grey region showing the central 95\% of the 
distribution of spawning biomass, and thin black lines 
showing three randomly selected simulation replicates. Catch is shown as grey 
bars in the historical period, which represent median catch in the projection, 
with thin vertical line segments showing the central 95\% of the catch 
distribution. Coloured circles on the right hand vertical axis show 
the biomass depletion level associated with the multi-species 
(closed blue circle) and single-species (open green circle) 
maximum sustainable yield.}\label{fig:fig6-tulipBtCtHMSPoor}
\end{figure}

\clearpage

\begin{figure}[htb]

{\centering \includegraphics[width=9in]{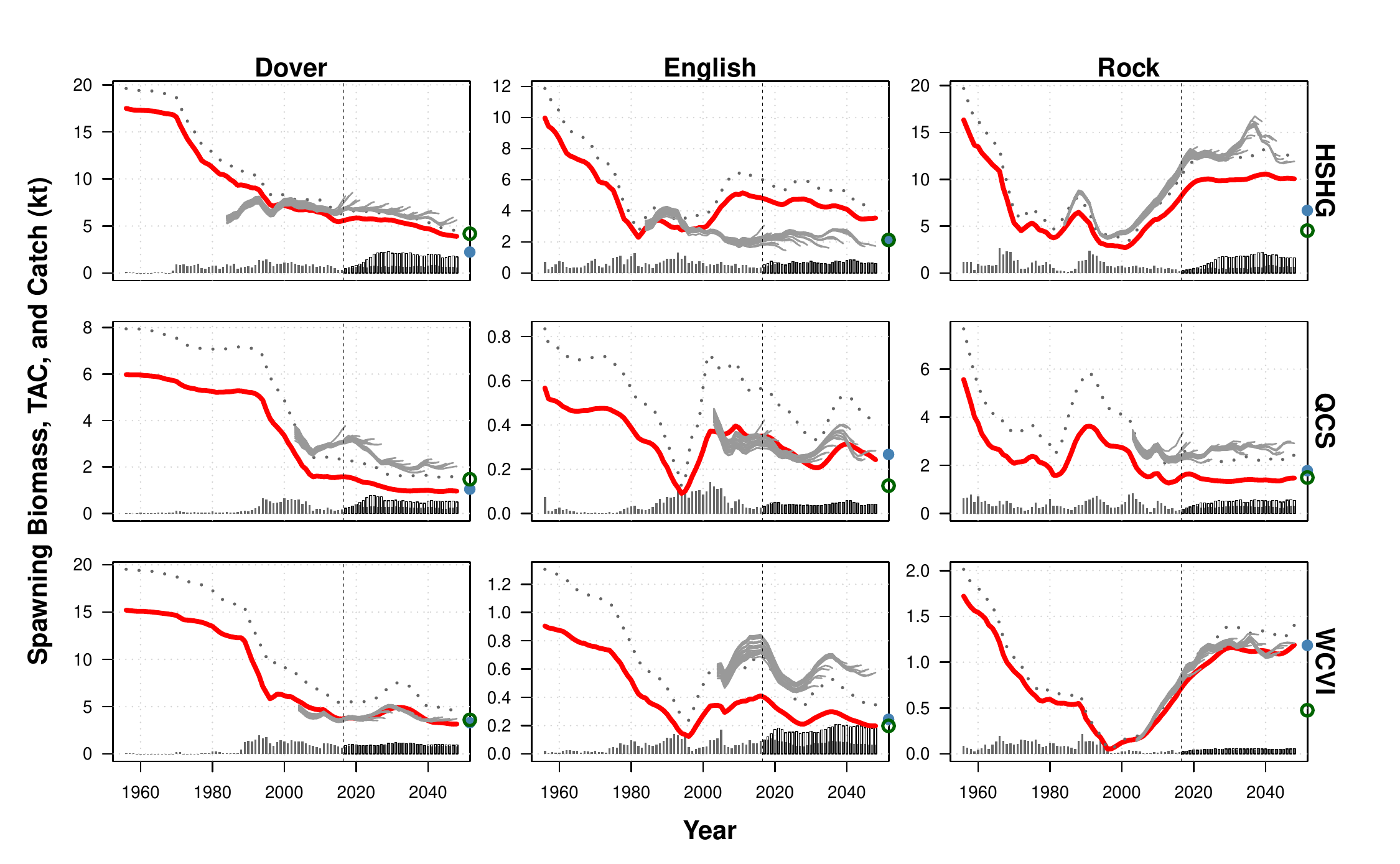} 

}

\caption{Operating model spawning stock
biomass (red line), commercial trawl vulnerable biomass (grey dotted line), 
retrospective assesment model estimates of spawning stock 
biomass (thin grey/purple lines), and catch and TACs (grey bars) 
from the first simulation replicate in the Poor data-quality scenario and
under the Hierarchical Multi-stock assessment model. Catch bars show realised 
catch in grey for the whole simulation period, and unfilled bars 
in the projection period show the difference between MP set TACs 
and realised catch. Coloured circles on the right hand vertical axis 
show the biomass level associated with the multi-species 
(closed blue circle) and single-species (open green circle) 
maximum sustainable yield.}\label{fig:fig7-retroBioSingleRepHMSPoor}
\end{figure}

\clearpage

\begin{figure}[htb]

{\centering \includegraphics[width=9in]{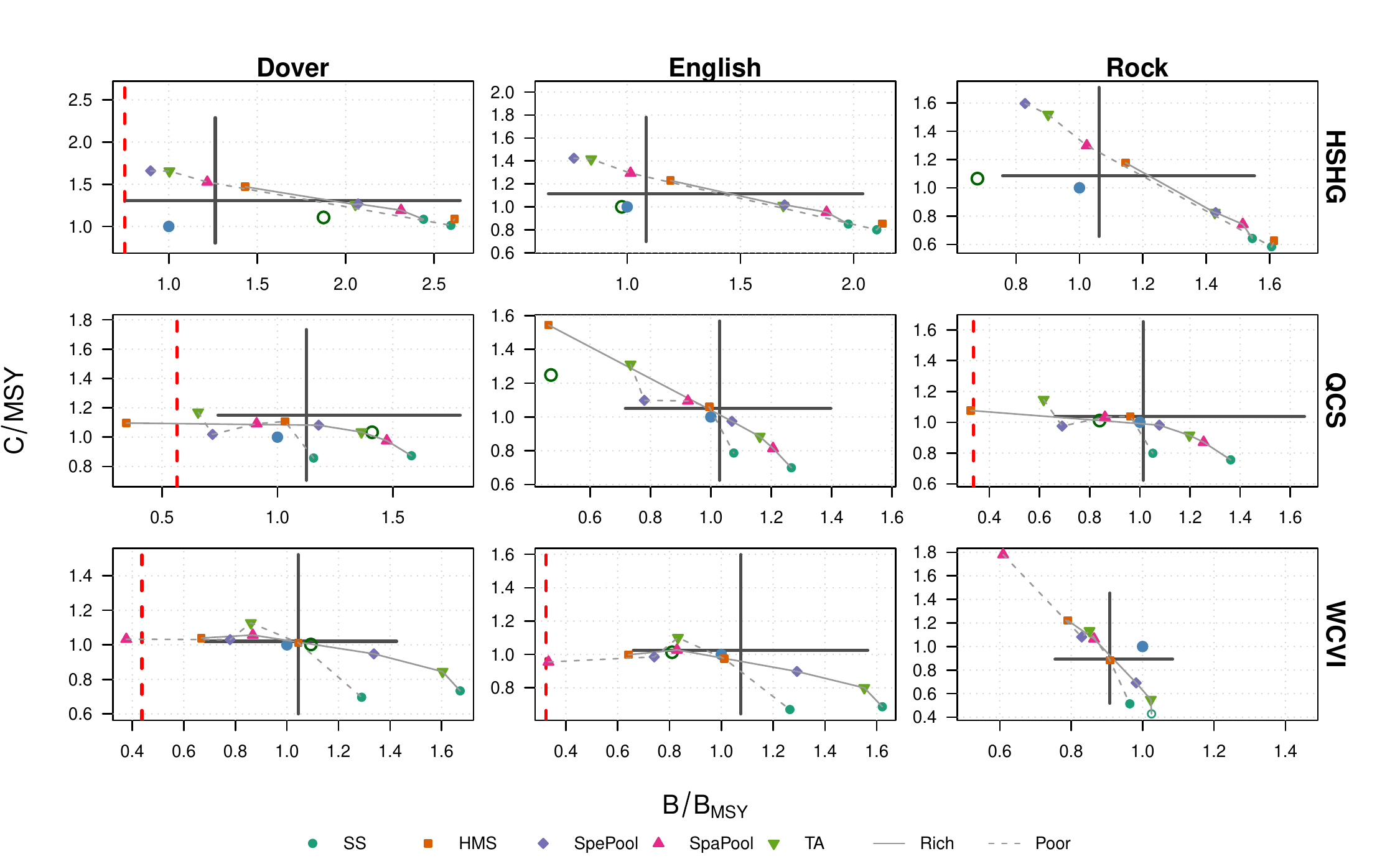} 

}

\caption{Tradeoff between catch and biomass 
during the 2028 - 2037 period implied by switching between 
different assessment models under Rich and Poor data-quality scenarios. 
Panels are gridded by species (columns) and stocks (rows), with biomass 
relative to $B_{MSY,MS,s,p}$ on the horizontal axis, and catch 
relative to $MSY_{MS,s,p}$ on the vertical axis. Distributions of
biomass and catch under the omniscient manager are shown by the 
black crosshair, with points indicating optimal biomass and yield
for single species maximum yield (open green circles) and 
multi-species maximum yield (closed blue circles). The biomass 
level at which a stock is critically overfished is shown as a vertical
red dashed line. Coloured point symbols show median biomass and catch 
for over all replicates for different assessment models, with assessment 
models under the same data-quality scenario joined by a solid line 
(Rich) or dashed line (Poor).}\label{fig:fig8-catchBioTradeoff}
\end{figure}

\clearpage

\begin{figure}[htb]

{\centering \includegraphics[width=8in]{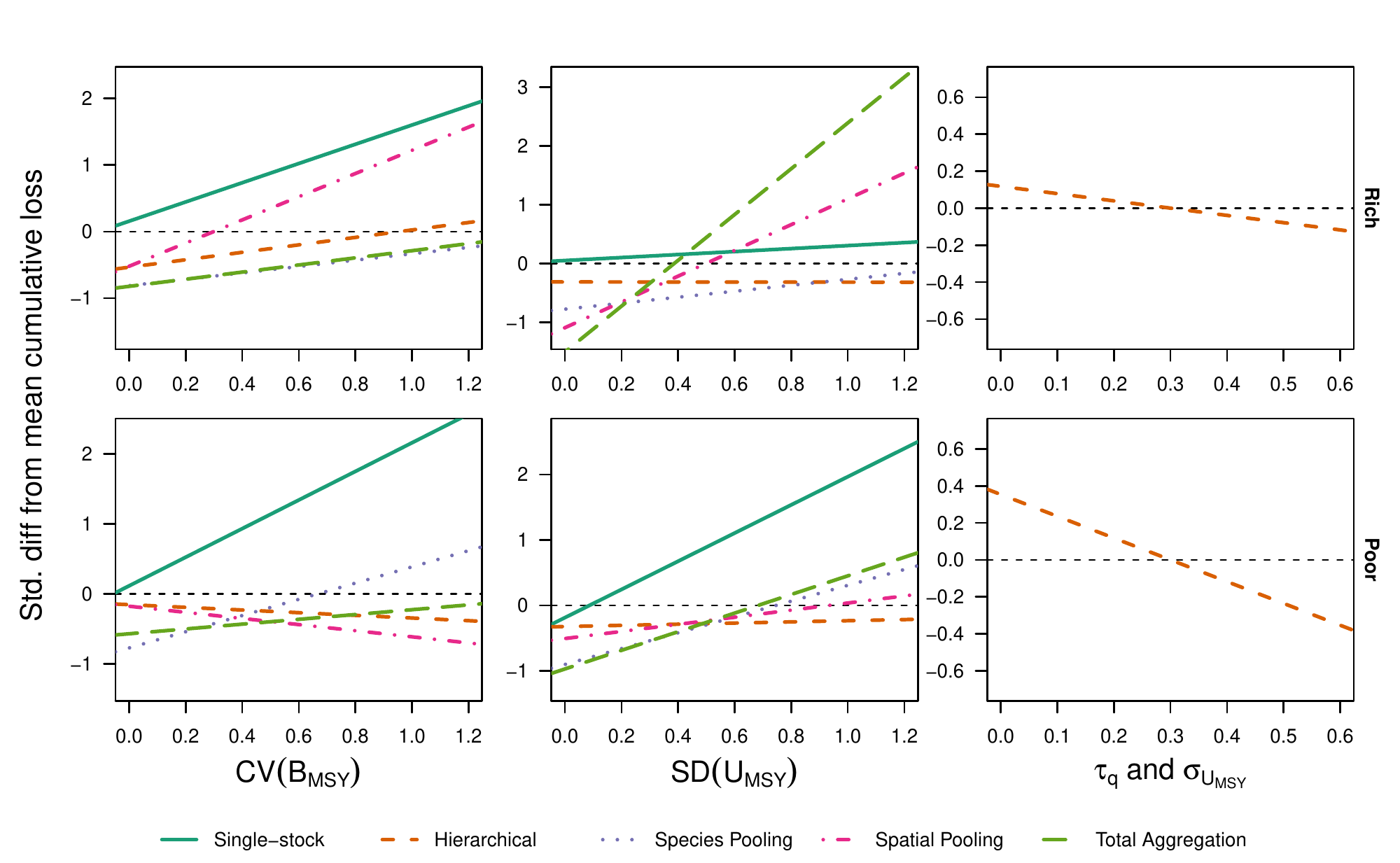} 

}

\caption{Regressions showing the average 
sensitivity of cumulative catch loss to the prior standard
deviations under each assessment model (colours, line types) for the 
data-rich (left column) and data-poor (right column) scenarios. 
The horizontal axis on each plot shows the prior standard deviation 
(CV for $B_{MSY}$), while the vertical axis shows the standardised 
difference between median cumulative loss for an assessment model and the mean
of median cumulative loss values over AMs, stratified by species
and area. See the online version of the journal for a full colour 
version of the plot.}\label{fig:fig9-regLossSensitivityRuns}
\end{figure}

\end{landscape}

\clearpage
\setcounter{table}{0}
\setcounter{figure}{0}

\hypertarget{appendix-appendices}{%
\appendix}

\hypertarget{the-operating-model}{%
\section{The operating model}\label{the-operating-model}}

The operating model was a standard age- and sex-structured
operating model, with additional structure for multi-species
and multi-stock population dynamics. DER complex species and
stocks were simulated assuming no ecological interactions or
movement between areas. The lack of movement may be unrealistic,
especially for Dover Sole given their extent, but this is how
the the DER complex stocks are currently managed in practice.
The lack of ecological interactions is more realistic for Dover
and English soles, as although both species are benthophagus,
there is evidence that they belong to different feeding guilds
\citep{pikitch1987use}.

DER complex abundance \(N_{a,x,s,p,t}\) for age \(a\), sex \(x\),
species \(s\) and stock \(p\) at the start of year \(t\) was
given by
\begin{equation*}
N_{a,x,s,p,t} = \left \{ \begin{array} {ll}
                          0.5 R_{s,p,t}  & a = 1, \\
                          N_{a-1,x,s,p,t-1} \cdot e^{-Z_{a-1,x,s,p,t-1}} & 1 < a < A, \\
                          N_{a-1,x,s,p,t-1} \cdot e^{-Z_{a-1,x,s,p,t-1}} + N_{a,x,s,p,t-1}\cdot e^{-Z_{a,x,s,p,t-1}} & a = A_s,
                          \end{array}
                \right.
\end{equation*}
where \(R_t\) is age-1 recruitment in year \(t\), \(Z_{a,x,s,p,t}\)
is the instantaneous total mortality rate, and \(A_s\) is the plus
group age for species \(s\).

Numbers-at-age were scaled to biomass-at-age by sex/species/area-
specific weight-at-age. Weight-at-age was an allometric function
of length-at-age
\begin{equation*}
w_{a,x,s,p} = \alpha_{x,s,p} \cdot L_{a,x,s,p}^{\beta_{x,s,p}}
\end{equation*}
where \(\alpha_{x,s,p}\) scaled between cm and kg, \(\beta_{x,s,p}\)
determined the rate of allometric growth, and \(L_{a,x,s,p}\) was
the length in cm of a fish of age \(a\), sex \(x\), species \(s\) and
stock \(p\). Length-at-age was given by the following Schnute
formulation of the von-Bertalanffy growth curve \citep{schnute1981versatile, francis2016estimating}
\begin{equation*}
L_{a} = \overline{L}_{A_1} - (\overline{L}_{A_2} - \overline{L}_{A_1}) \cdot 
            \left( \frac{e^{-k A_1} - e^{-k a} )} {e^{-k A_1} - e^{-k A_2} } \right) 
\end{equation*}
where \(A_1\) and \(A_2\) are well spaced reference ages,
\(\overline{L}_{A_1}\) and \(\overline{L}_{A_2}\) are the mean lengths
in cm of fish at ages \(A_1\) and \(A_2\), and \(k\) is the
growth coefficient. Note that in the growth model we dropped the sex,
species and stock subscripts for concision.

The maturity-at-age ogive was modelled as a logistic function
\begin{equation*}
m_{a,s,p} = \left( 1 + e^{-\frac{\ln 19 (a - a_{50,s,p}^{mat})}{a_{95,s,p}^{mat} - a_{50,x,s,p}^{mat}}}\right)^{-1},
\end{equation*}
where \(m_{a,s,p}\) was the proportion of age-\(a\) female fish
of species \(s\) in stock \(p\) that were mature, and \(a_{50,s,p}^{mat}\)
and \(a_{95,s,p}^{mat}\) are the ages at which 50\% and 95\% of fish
of age-\(a\), species \(s\) and stock \(p\) were mature.

Female spawning stock biomass was calculated as
\begin{equation*}
B_{s,p,t} = \sum_{a} N_{a,x',s,p,t} m_{a,s,p} w_{a,x',s,p},
\end{equation*}
where \(x'\) denotes female fish only. Spawning stock biomass
was used to calculate expected Beverton-Holt recruitment, which
then had recruitment process errors applied
\begin{equation*}
R_{s,p,t+1} = \frac{R_{s,p,0} \cdot 4h_{s,p} \cdot B_{s,p,t} }
                    { B_{s,p,0} \cdot (1 - h_{s,p} ) + (5h_{s,p} - 1) \cdot B_{s,p,t}  }
                    \cdot e^{\epsilon_{s,p,t+1} - 0.5\sigma_{R,s,p}^2},
\end{equation*}
where \(R_{s,p,0}\) is unfished equilibrium recruitment, \(B_{s,p,t}\)
is the spawning stock biomass at time \(t\), \(B_{s,p,0}\) is unfished spawning
stock biomasss, \(h_{s,p}\) is stock-recruit steepness (average proportion
of \(R_{s,p,0}\) produced when \(B_{s,p,t} = .2B_{s,p,0}\)), and
\(\epsilon_{s,p,t}\) is the recruitment process error with standard
deviation \(\sigma_{R,s,p}\).

The operating model was initialised in 1956 at unfished equilibrium
for all species \(s\) and areas \(p\), with numbers-at-age in 1956
given by
\begin{equation*}
N_{a,x,s,p,1956} = \left \{ \begin{array} {ll}
                          0.5 R_{s,p,0}  & a = 1, \\
                          N_{a-1,x,s,p,1956} \cdot e^{-M_{x,s,p}} & 1 < a < A, \\
                          N_{a-1,x,s,p,1956} \cdot \frac{e^{-M_{x,s,p}}}{1 - e^{-M_{x,s,p}}} & a = A,
                          \end{array}
                \right.
\end{equation*}

Fishery removals were assumed to be continuous throughout the year,
with fishing mortality-at-age
\begin{equation*}
F_{a,x,s,p,f,t} = S_{a,x,s,p,f} \cdot F_{s,p,f,t},
\end{equation*}
where \(F_{s,p,f,t}\) is the fully selected fishing mortality rate for
fleet \(f\) at time \(t\), and \(S_{a,x,s,p,f}\) is the selectivity-at-age
\(a\) for sex \(x\) in species \(s\) and area \(p\) by fleet \(f\). Selectivity-at
-age was modeled as a logistic function of length-at-age
\[
S_a = \left( 1 + \exp \left( \frac{-\ln 19 (L_a - l^{sel}_{50})}{l^{sel}_{95} - l^{sel}_{50}}  \right)   \right)^{-1},
\]
where \(L_a\) is length-at-age, defined above, and \(l^{sel}_{50}\)
and \(l^{sel}_{95}\) are the length-at-50\% and length-at-95\% selectivity,
respectively; stock, species and fleet subscripts are left off for concision.
Catch-at-age was then found via the Baranov catch equation
\begin{equation*}
C_{a,x,s,p,f,t} = (1 - e^{-Z_{a,x,s,p,f,t}}) \cdot N_{a,x,s,p,t} w_{a,x,s,p} \frac{F_{a,x,s,p,f,t}}{Z_{a,x,s,p,f,t}},
\end{equation*}
where total mortality-at-age is defined as
\[
Z_{a,x,s,p,f,t} = M_{x,s,p} + S_{a,x,s,p,f} \cdot F_{a,x,s,p,f,t}.
\]

\hypertarget{observation-error-standard-deviations}{%
\subsection{Observation error standard deviations}\label{observation-error-standard-deviations}}

Operating model observation error standard deviations were derived
from estimates from fitting a hierarchical age-structured model to
DER complex data {[}Johnson and Cox, in prep{]}. To improve convergence
in the simulations, we multipled all estimates by \(0.3\) to improve
observation model precision (multiplier found by trial and error).

\begin{table}[!h]

\caption{\label{tab:obsErrTab}Log-normal observation error standard deviations for all DER complex
biomass indices}
\centering
\begin{tabular}[t]{lllll}
\toprule
\multicolumn{1}{c}{ } & \multicolumn{4}{c}{Observation Error SD} \\
\cmidrule(l{3pt}r{3pt}){2-5}
Stock & Historical & Modern & HS Ass. & Syn\\
\midrule
\addlinespace[0.3em]
\multicolumn{5}{l}{\textbf{Dover sole}}\\
\hspace{1em}HSHG & 0.174 & 0.158 & 0.328 & 0.138\\
\hspace{1em}QCS & 0.187 & 0.155 &  & 0.149\\
\hspace{1em}WCVI & 0.221 & 0.161 &  & 0.092\\
\addlinespace[0.3em]
\multicolumn{5}{l}{\textbf{English sole}}\\
\hspace{1em}HSHG & 0.170 & 0.163 & 0.255 & 0.191\\
\hspace{1em}QCS & 0.225 & 0.171 &  & 0.203\\
\hspace{1em}WCVI & 0.216 & 0.174 &  & 0.136\\
\addlinespace[0.3em]
\multicolumn{5}{l}{\textbf{Rock sole}}\\
\hspace{1em}HSHG & 0.164 & 0.161 & 0.254 & 0.187\\
\hspace{1em}QCS & 0.176 & 0.184 &  & 0.205\\
\hspace{1em}WCVI & 0.198 & 0.233 &  & 0.214\\
\bottomrule
\end{tabular}
\end{table}

\clearpage

\hypertarget{omniscient-manager-optimisation}{%
\section{Omniscient Manager Optimisation}\label{omniscient-manager-optimisation}}

We defined penalty functions so that inside their respective
desired regions the penalty was zero, and otherwise the penalty grew
as a cubic function of distance from the desired region. For example,
a penalty designed to keep a measurement \(x\) above a the desired
region boundary \(\epsilon\) is of the form
\begin{equation}
\mathcal{P}(x , \epsilon) = \left\{
  \begin{array}{ll}
    0 & x \geq \epsilon, \\
    |x - \epsilon|^3 & x < epsilon. \\
  \end{array} \right. 
\end{equation}
This form has a several advantages over simple linear penalties,
or a logarithmic barrier penalty \citep{srinivasan2008tracking}. First,
the cubic softens the boundary threshold \(\epsilon\), effectively
allowing a crossover if doing so favours another portion of the
objective function. Second, unlike lower degree polynomials, cubic
functions remain closer to the \(x\)-axis when \(|x-\epsilon| < 1\).
Third, zero penalty within in the desirable region stops
the objective function from favouring regions far from the boundaries
of penalty functions. In contrast, a logarithmic function would
favour overly conservative effort series to keep biomass far from
a lower depletion boundary. Finally, the cubic penalty function and its
first two derivatives are continuous at every point \(x\), allowing for
fast derivative-based optimisation methods.

We used a cubic spline of effort in each area to reduce
the number of free parameters in the optimisation. For each
area, 9 knot points were distributed across the full 40 year
projection, making them spaced by 5 years. We padded
the omniscient manager simulations by an extra eight
years over the stochastic simulations to avoid any
possible end effects of the spline entering the
performance metric calculations. Effort splines
were constrained to be between 0 and 120 times the
operating model \(E_{MSY,p}\), by replacing any value
outside that range with the closest
value inside the range (i.e.~negative values by \emph{zero},
large values by \(120 E_{MSY_p}\)).

\clearpage

\hypertarget{hierarchical-model-performance-under-relaxed-shrinkage-prior-sds}{%
\section{Hierarchical model performance under relaxed shrinkage prior SDs}\label{hierarchical-model-performance-under-relaxed-shrinkage-prior-sds}}

\begin{landscape}
\begin{figure}[htb]

{\centering \includegraphics[width=9in]{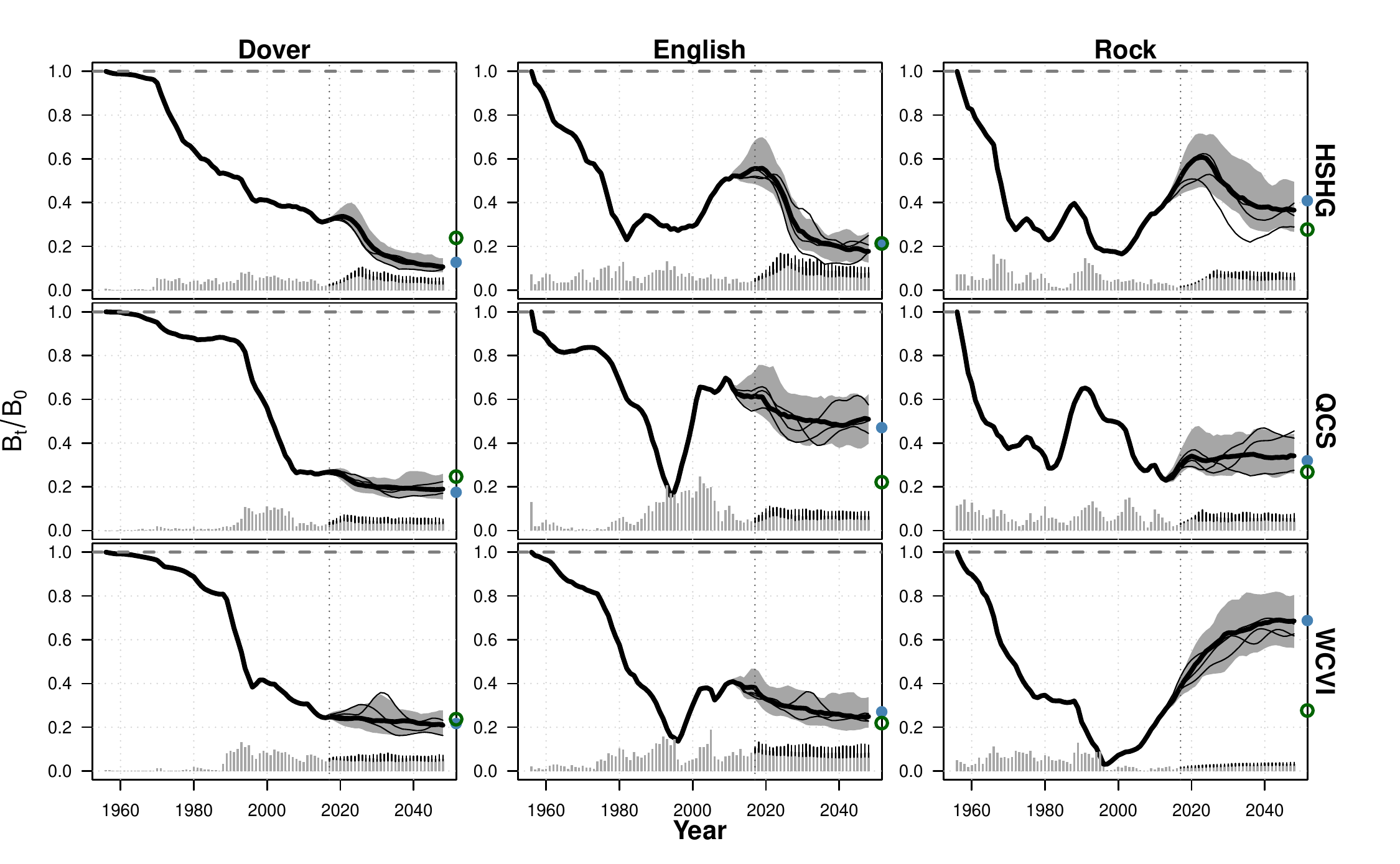} 

}

\caption{Spawning biomass depletion and relative catch 
simulation envelopes for all nine DER complex management units when 
assessed by the hierarchical multi-stock asssessment model under the 
Poor OM data quality scenario, when hierarchical shrinkage prior SDs
were $\sigma_{U_{MSY}} = \tau_{q} = 0.5$. Median biomass is shown by 
the thick black line, with the grey region showing the central 95\% of the 
distribution of spawning biomass, and thin black lines 
showing three randomly selected simulation replicates. Catch is shown as grey 
bars in the historical period, which represent median catch in the projection, 
with thin vertical line segments showing the central 95\% of the catch 
distribution. Coloured circles on the right hand vertical axis show 
the biomass depletion level associated with the multi-species 
(closed blue circle) and single-species (open green circle) 
maximum sustainable yield.}\label{fig:figC1-tulipBtCtHMSPoorHierSD}
\end{figure}

\clearpage

\begin{figure}[htb]

{\centering \includegraphics[width=9in]{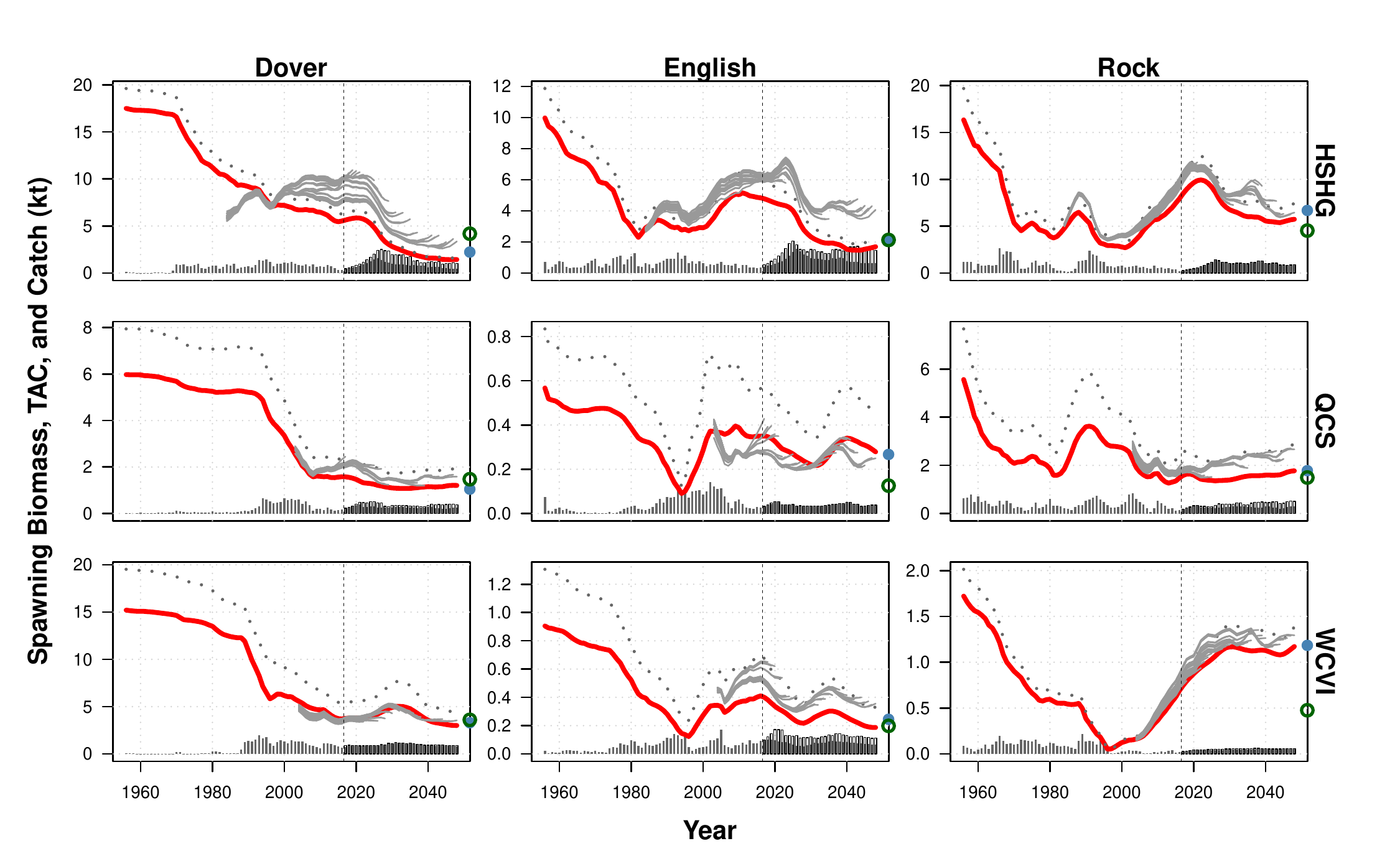} 

}

\caption{Operating model spawning stock
biomass (red line), commercial trawl vulnerable biomass (grey dotted line), 
retrospective assesment model estimates of spawning stock 
biomass (thin grey/purple lines), and catch and TACs (grey bars) 
from the first simulation replicate in the Poor data quality scenario
under the hierarchical multi-stock assessment model when hierarchical 
shrinkage prior SDs were $\sigma_{U_{MSY}} = \tau_{q} = 0.5$. Catch 
bars show realised catch in grey for the whole simulation period, and 
unfilled bars in the projection period show the difference between MP 
set TACs and realised catch. Coloured circles on the right hand vertical axis 
show the biomass level associated with the multi-species 
(closed blue circle) and single-species (open green circle) 
maximum sustainable yield.}\label{fig:figC2-retroBioSingleRepHMSPoorHierSD}
\end{figure}

\end{landscape}

\bibliography{bib/library.bib}

\end{document}